\documentclass[11pt,a4paper]{article}
\usepackage[margin=25mm]{geometry}
\usepackage{amsmath,amssymb,amsthm,graphicx,booktabs,array,tabularx}
\usepackage[numbers,sort&compress]{natbib}
\usepackage{microtype,enumitem,placeins}
\usepackage[colorlinks=true,allcolors=blue]{hyperref}
\newcommand{\R}{\mathbb R}
\newcommand{\g}{\boldsymbol g}
\newcommand{\norm}[1]{\left\|#1\right\|}
\newcommand{\Cov}{\operatorname{Cov}}
\newcommand{\RSS}{\operatorname{RSS}}
\newtheorem{proposition}{Proposition}

\title{Reconstructing constant point heat sources from few temperature sensors: sparse selection and spatial resolution}
\author{Xiaomei Yang\thanks{School of Mathematics, Southwest Jiaotong University. Email: yangxiaomath@swjtu.edu.cn}\and 
Zhiliang Deng\thanks{Corresponding author: dengzhl@uestc.edu.cn. School of Mathematical Science, University of Electronic Science and Technology of China.}}
\date{}

\begin{document}
\maketitle
\begin{abstract}
We investigate the recovery of an unknown number of stationary point heat sources, together with their locations and constant positive strengths, from temperature histories recorded by a few sensors. The reconstruction combines candidate selection by orthogonal matching pursuit or least angle regression with continuous constrained fitting. A noise-based discrepancy rule selects the smallest accepted source count among those tested. The analysis distinguishes source-count selection, individual-parameter recovery, and aggregate estimation. For a close pair, the local single-source residual is quadratic in separation when the projected curvature is nonzero, whereas a compensating unit parameter direction produces a linearized response bounded by a constant times the cube of separation. This difference explains why a pair may be distinguishable from a single source while its individual parameters remain poorly determined. A conditional stability bound controls errors in total strength and the strength-weighted center, and a local Gaussian approximation relates single-source rejection to noise and separation. Numerical experiments assess sensor configuration, temporal sampling, noise, separation, and initialization, with additional tests using independently generated finite-plate data and a three-dimensional model. Correct source-count selection need not imply accurate individual parameters, whereas unresolved pairs can retain accurate aggregate estimates.
\end{abstract}

\section{Introduction}

Identifying localized heat sources from temperature measurements is important in a wide range of engineering applications, such as thermal monitoring of conducting bodies \cite{Lefevre2002,LeNiliot2004}, hot-spot detection in electronics packages \cite{Krane2022}, and heat-source estimation in machine-tool spindles \cite{Than2017}. In practice, access and placement constraints often limit the number of available sensors. Thermal diffusion further complicates reconstruction by smoothing the temperature signatures of localized sources, with recoverability depending on sensor configuration, sampling times, and measurement noise. When the source count is unknown, the inverse problem requires determining how many sources are supported by the measurements, together with their locations and strengths. These challenges make the reconstruction computationally demanding. To address these difficulties, various methods have been developed for inverse heat-source identification, including parameter estimation, regularization, and sparse reconstruction. Ling and co-workers \cite{Ling2006,Ling2009} developed methods based on fundamental solutions, while Le Niliot and Lefèvre \cite{LeNiliot2004} combined a boundary element formulation with nonlinear estimation of source locations and strengths. Sparse formulations are particularly relevant when only a small number of candidate locations are expected to contain active sources. For the related problem of recovering an initial temperature distribution, Li et al.\ \cite{Li2014} represented the unknown distribution by sparse Dirac masses and used $\ell_1$-constrained minimization. Lu et al.\ \cite{Lu2019} developed finite-element-based sparse regularization for inverse heat-source identification. More recently, Yan and Lu \cite{Yan2025} proposed a decomposed physics-based compressive-sensing framework for reconstruction from sparse temperature measurements under uncertain boundary conditions.

Source identification has also been studied using sparse boundary flux measurements. Rundell and Zhang \cite{Rundell2020} established uniqueness, up to a multiplicative ambiguity, for sources with separable spatial and temporal factors under suitable assumptions, and proposed an alternating regularization scheme for reconstruction. Gu et al.\ \cite{Gu2025} considered the identification of a single point source by minimizing a least-squares objective using gradient descent. Gong et al.\ \cite{Gong2026} established uniqueness results for the location and time-dependent amplitude of a single point source and presented an alternating reconstruction scheme for these unknowns.

In this work, we consider an unknown number of stationary point heat sources with constant positive strengths and seek to recover their number, locations, and strengths from temperature histories recorded by a few sensors. Beyond fitting the measured temperatures, we examine which source properties can be reliably recovered. A small fitting residual alone does not ensure accurate recovery of the underlying sources. For example, a close group of sources may produce nearly the same temperature response as a single source with their combined strength. We therefore distinguish source-count selection, recovery of individual locations and strengths, and estimation of aggregate quantities, namely total strength and the strength-weighted center.

The reconstruction uses a dictionary of candidate unit-source responses. Orthogonal matching pursuit (OMP) \cite{Pati1993} or least angle regression (LARS) \cite{Efron2004} selects initial locations for continuous constrained fitting. A noise-based discrepancy rule selects the smallest accepted model size among those tested, and the final locations need not lie on the candidate grid. This procedure builds on established sparse-selection and continuous-refinement ideas \cite{Mamandipoor2016,Boyd2017}. Related two-stage and measure-based methods also address sparse source reconstruction \cite{Biccari2023,Mattuschka2026}; the latter includes unknown source counts and continuously operating sources. Our analysis connects this reconstruction procedure with localization sensitivity, local single-source rejection, and aggregate recovery. In particular, a quadratic single-source residual and a cubic upper bound along a two-source parameter direction distinguish rejection of a simpler model from sensitivity of individual source parameters. A conditional stability estimate addresses aggregate parameters, and a local Gaussian limit supplies a noise-dependent reference for single-source rejection. Numerical experiments assess these results and examine the roles of observation configuration, noise, separation, and initialization.

The remainder of the paper is organized as follows. Section~\ref{sec:model} introduces the thermal response and observation model. Section~\ref{sec:method} describes sparse selection, continuous fitting, and source-count selection. Section~\ref{sec:resolution} analyzes source sensitivity and close-source resolution. Section~\ref{sec:numerics} presents the numerical assessment, and Section~\ref{sec:conclusions} discusses the findings and concludes the paper.

\section{Thermal responses and the inverse problem}\label{sec:model}

\subsection{Constant sources and temperature observations}

Let \(u\) denote the temperature increment relative to a known source-free baseline. For a known thermal diffusivity \(\kappa>0\), consider
\begin{equation}\label{eq:pde}
\partial_tu-\kappa\Delta u=\sum_{j=1}^{K}a_j\delta_{z_j},\qquad u(\cdot,0)=0,
\end{equation}
in \(\mathcal D\subset\R^d\), with \(d=2\) or \(3\), and with prescribed boundary conditions when \(\mathcal D\) is bounded. 
The sources remain at fixed locations and operate at constant strengths from $t=0$, with unknown count $K$, locations $z_j\in\mathcal Z$, and strengths $a_j>0$. 
The admissible region \(\mathcal Z\subset\mathcal D\) is compact and separated from the sensors. Distinct physical sources are represented by distinct locations; coincident sources are combined into one source. Here \(\delta_z\) denotes a unit Dirac point mass at \(z\). The strength is the heat-release rate divided by the volumetric heat capacity, or by the corresponding areal heat capacity in a two-dimensional plate model.

After baseline subtraction, assume homogeneous, time-independent linear boundary conditions when the domain is bounded. Let \(G_{\mathcal D}(b,z,s)\) denote the temperature response at \(b\), at elapsed time \(s\), to a unit instantaneous point source at \(z\). The response to a unit-strength source operating continuously from \(t=0\) is then
\begin{equation}\label{eq:response}
R_{\mathcal D}(b,z,t)=\int_0^tG_{\mathcal D}(b,z,s)\,ds.
\end{equation}
At sensor locations \(b_i\), \(i=1,\ldots,S\), and positive times \(t_\ell\), \(\ell=1,\ldots,T\), define \(M=ST\) and
\begin{equation}\label{eq:data}
\g(z)=\big(R_{\mathcal D}(b_i,z,t_\ell)\big)_{i,\ell},\qquad
F_K(\theta)=\sum_{j=1}^Ka_j\g(z_j),\qquad y=F_{K_*}(\theta_*)+e.
\end{equation}
Here \(\theta=(z_1,a_1,\ldots,z_K,a_K)\in\R^{(d+1)K}\), and a subscript \(*\) denotes the true count or the true parameters. The vector \(\g(z)\in\R^M\) contains the 
responses of a unit-strength source at all sampled sensor-time pairs.  All sampling times for each sensor are grouped together in the observation vector. 
We assume additive Gaussian measurement noise,
\[
 e\sim\mathcal N(0,\sigma^2I_M),
\]
where $\sigma>0$ is known and $I_M$ is the $M\times M$
identity matrix. Thus, measurement errors are independent
and have a common variance $\sigma^2$.
The inverse problem is to recover the source count, locations,
and strengths from the temperature observation vector $y$.
\subsection{Whole-space response}

In \(\mathcal D=\R^d\), write \(r=|b-z|>0\). The response is
\begin{equation}\label{eq:whole}
R_{\R^d}(b,z,t)=H_d(r,t)=\int_0^t(4\pi\kappa s)^{-d/2}\exp\!\left(-\frac{r^2}{4\kappa s}\right)\,ds,
\end{equation}
with explicit forms
\begin{equation}\label{eq:whole_explicit}
H_2(r,t)=\frac{1}{4\pi\kappa}E_1\!\left(\frac{r^2}{4\kappa t}\right),\qquad
H_3(r,t)=\frac{1}{4\pi\kappa r}\operatorname{erfc}\!\left(\frac{r}{2\sqrt{\kappa t}}\right).
\end{equation}
Here \(E_1(x)=\int_x^\infty e^{-s}s^{-1}\,ds\) and \(\operatorname{erfc}(x)=2\pi^{-1/2}\int_x^\infty e^{-s^2}\,ds\).   
The positive distance between the sensors and the admissible source
region excludes singular readings at \(r=0\). The ratio
\(r/\sqrt{\kappa t}\) compares the source--sensor distance with the
diffusion length and helps explain the dependence of the response
on distance and sampling time.

For a bounded domain satisfying the boundary assumptions above, the reconstruction uses its unit-source response $R_{\mathcal D}$ in place of the whole-space kernel. The finite-plate implementation is specified in Section~\ref{sec:transfer}.

\section{Reconstruction by sparse selection and continuous fitting}
\label{sec:method}

The reconstruction begins with a grid of candidate source locations. OMP or LARS selects an ordered subset of their temperature responses. For each tested source count $k$, the first $k$ candidates initialize a continuous fit of locations and nonnegative strengths. Counts are tested in increasing order, and the first converged fit within the noise threshold is accepted.

Throughout, $|\cdot|$ denotes the Euclidean norm in spatial or
parameter coordinates, $\norm{\cdot}_2$ denotes the Euclidean
norm of observation vectors, and $\norm{\cdot}_{\rm op}$
denotes the induced Euclidean operator norm, while \(\norm{\cdot}_{\rm F}\)
denotes the Frobenius norm of a matrix.
We write \(D\g(z)\in\R^{M\times d}\) for the location Jacobian
and \(D^r\g(z)[v_1,\ldots,v_r]\) for its derivative of order
\(r\) applied to the indicated directions.
Jacobian sensitivity measures and parameter-error bounds use the stated
location and strength coordinates and depend on their scaling.

\subsection{Thermal response dictionary and sparse representation}

Choose a candidate grid
\(\Gamma=\{\xi_n\}_{n=1}^N\subset\mathcal Z\).
For each candidate location \(\xi_n\), evaluate the temperature history generated by a unit constant source at every sensor and sampling time. These responses form the dictionary
\begin{equation}\label{eq:dictionary}
 \mathsf A_{:,n}=\g(\xi_n),\qquad
 d_n=\norm{\mathsf A_{:,n}}_2,\qquad
 X_{:,n}=\frac{\mathsf A_{:,n}}{d_n}.
\end{equation}
Only candidates with \(d_n>0\) are retained.
Writing \(x_n=X_{:,n}\), every normalized column satisfies
\(\norm{x_n}_2=1\).

If the sources were restricted to the candidate grid, their temperature data would admit the representation
\begin{equation}\label{eq:grid_regression}
 y\approx\mathsf A\alpha=X\beta,\qquad \beta_n=d_n\alpha_n,
\end{equation}
where \(\alpha\in\R^N\) denotes the grid-source strengths and only a small number of entries are nonzero. These strengths are distinct from the strengths \(a_j\) of the continuously located physical sources. The sparse selection stage searches for columns of \(X\) that explain the measured temperature vector. Column normalization allows this search to compare response patterns without automatically favoring candidates with larger response norms. Since the known source-free baseline has been removed, neither the columns nor the data are centered, and no intercept is fitted.

OMP and LARS use signed regression coefficients to select candidate
responses from the dictionary. The selected locations initialize a
constrained fit of continuous source positions and nonnegative strengths
using the unnormalized response. The grid therefore provides a discrete
representation for candidate selection, while the final source
parameters are estimated continuously. The following bound quantifies
the response error introduced by approximating continuous source
locations with nearby grid points.

\begin{proposition}
\label{prop:grid}
Suppose that \(\g\) is continuously differentiable on a neighborhood of a convex region separated from the sensors and containing each source \(z_j\) and its nearest candidate \(\xi_j\), with \(\norm{D\g(z)}_{\rm op}\leq C_1\) on that region. If \(|z_j-\xi_j|\leq h_\Gamma\), then
\begin{equation}\label{eq:grid}
 \norm{\sum_ja_j\g(z_j)-\sum_ja_j\g(\xi_j)}_2
 \leq C_1h_\Gamma\sum_ja_j.
\end{equation}
For a Cartesian grid with coordinate spacing \(h\), \(h_\Gamma\leq\sqrt d\,h/2\), provided every source lies in a grid cell whose vertices are retained in \(\Gamma\).
\end{proposition}
\begin{proof}
For each source, the segment
\(\gamma_j(s)=z_j+s(\xi_j-z_j)\), \(0\leq s\leq1\), remains
in the convex region where the derivative bound holds.
The fundamental theorem of calculus implies
\[
 \g(\xi_j)-\g(z_j)
 =\int_0^1D\g(\gamma_j(s))(\xi_j-z_j)\,ds.
\]
Consequently,
\(\norm{\g(\xi_j)-\g(z_j)}_2\leq C_1|\xi_j-z_j|\leq C_1h_\Gamma\).
Multiplication by \(a_j>0\), summation, and the triangle
inequality establish \eqref{eq:grid}. Coincident grid assignments
are combined by adding their strengths. Every point in a Cartesian
cell lies within \(\sqrt d\,h/2\) of a nearest vertex.
\end{proof}
The approximation bound does not impose a strength cap at a grid node. If several sources are assigned to the same node, their combined strength may exceed the upper bound used for a single fitted source.

\subsection{Sparse candidate selection}
\subsubsection*{Orthogonal matching pursuit}

Let \(B\) be the maximum number of candidates retained for physical fitting. OMP constructs the candidate list by repeatedly selecting the response most strongly correlated with the unexplained part of the data \cite{Pati1993}. Initialize the selected index set, the fitted data, and the residual as
\[
 \mathcal A_0=\varnothing,\qquad
 \mu^{(0)}=0,\qquad
 \boldsymbol\rho^{(0)}=y.
\]
At iteration \(m\), perform the following steps.

\paragraph{Step 1: select a candidate location.}
For every unselected candidate, compute
\[
 c_n^{(m)}=x_n^T\boldsymbol\rho^{(m-1)},\qquad n\notin\mathcal A_{m-1}.
\]
Select
\begin{equation}\label{eq:omp_select}
 n_m\in\operatorname*{arg\,max}_{n\notin\mathcal A_{m-1}}
 \left|c_n^{(m)}\right|,
 \qquad
 \mathcal A_m=\mathcal A_{m-1}\cup\{n_m\}.
\end{equation}
Thus, \(\xi_{n_m}\) is the candidate whose normalized temperature response has the largest absolute correlation with the current residual.

\paragraph{Step 2: refit all selected responses.}
Let \(X_{\mathcal A_m}\) contain the selected columns. Compute the signed least-squares coefficients
\begin{equation}\label{eq:omp_refit}
 \widehat\beta_{\mathcal A_m}
 \in\operatorname*{arg\,min}_{b\in\R^m}
 \norm{X_{\mathcal A_m}b-y}_2^2.
\end{equation}
The refit updates every selected coefficient when a new candidate enters.

\paragraph{Step 3: update the residual.}
Set
\begin{equation}\label{eq:omp_residual}
 \mu^{(m)}=X_{\mathcal A_m}\widehat\beta_{\mathcal A_m},
 \qquad
 \boldsymbol\rho^{(m)}=y-\mu^{(m)}.
\end{equation}
For an exact least-squares solve,
\(X_{\mathcal A_m}^T\boldsymbol\rho^{(m)}=0\).
The residual is therefore orthogonal to the selected response span.

The iterations stop after \(\min(B,N)\) selections, or earlier when the remaining candidates satisfy
\[
 \max_{n\notin\mathcal A_m}|x_n^T\boldsymbol\rho^{(m)}|
 \leq \tau_{\rm corr}\norm y_2.
\]
Here \(\tau_{\rm corr}>0\) is a numerical correlation tolerance specified in Section~\ref{sec:protocol}. The output used by the reconstruction stage is the selection order \((n_1,\ldots,n_{N_{\rm sel}})\), with \(N_{\rm sel}\leq B\), rather than the signed dictionary coefficients.

\subsubsection*{Least angle regression}

LARS also starts with the candidate having the largest absolute correlation with the data. Its subsequent updates differ from OMP: the fitted data move along a direction that is equally correlated with all signed active columns, until another candidate reaches the same absolute residual correlation \cite{Efron2004}. We use the pure least-angle regression path.

Initialize
\[
 \beta=0,\qquad \mu=0,\qquad \boldsymbol\rho=y,\qquad c_n=x_n^T\boldsymbol\rho.
\]
Choose an index attaining \(\max_n|c_n|\) and place it in the active set \(\mathcal A\). Record its index as the first entry of the candidate list. The following steps describe a path segment with a linearly independent active set.

\paragraph{Step 1: construct the equiangular direction.}
At a point where a new variable enters the LARS path, the active columns have a common absolute correlation
\[
 C_{\rm corr}=\max_n|c_n|,\qquad |c_j|=C_{\rm corr},\quad j\in\mathcal A.
\]
For each active index, define its correlation sign \(s_j=\operatorname{sign}(c_j)\) and form
\[
 \widetilde X_{\mathcal A}=[s_jx_j]_{j\in\mathcal A},\qquad
 \mathsf G_{\mathcal A}=\widetilde X_{\mathcal A}^T\widetilde X_{\mathcal A}.
\]
Let \(\boldsymbol1_{\mathcal A}\) denote the vector of ones of length \(|\mathcal A|\). Define
\begin{equation}\label{eq:lars_direction}
 \omega_{\mathcal A}
 =\left(
 \boldsymbol1_{\mathcal A}^T\mathsf G_{\mathcal A}^{-1}
 \boldsymbol1_{\mathcal A}
 \right)^{-1/2},
 \qquad
 w_{\mathcal A}
 =\omega_{\mathcal A}\mathsf G_{\mathcal A}^{-1}
 \boldsymbol1_{\mathcal A},
 \qquad
 u_{\mathcal A}=\widetilde X_{\mathcal A}w_{\mathcal A}.
\end{equation}
This direction satisfies
\[
 \norm{u_{\mathcal A}}_2=1,\qquad
 \widetilde X_{\mathcal A}^Tu_{\mathcal A}
 =\omega_{\mathcal A}\boldsymbol1_{\mathcal A}.
\]
Consequently, moving along \(u_{\mathcal A}\) reduces the signed residual correlations of all active columns at the same rate. The expressions in \eqref{eq:lars_direction} are evaluated by solving linear systems rather than forming an explicit inverse.

\paragraph{Step 2: determine when another candidate enters.}
For each dictionary column, compute
\[
 v_n=x_n^Tu_{\mathcal A}.
\]
If the fitted data move by a distance \(\gamma\) along \(u_{\mathcal A}\), then
\[
 \mu(\gamma)=\mu+\gamma u_{\mathcal A},\qquad
 c_n(\gamma)=c_n-\gamma v_n.
\]
For an active column,
\[
 s_jc_j(\gamma)=C_{\rm corr}-\gamma\omega_{\mathcal A}.
\]
An inactive candidate enters when its absolute correlation first reaches this common active value:
\[
 |c_n-\gamma v_n|=C_{\rm corr}-\gamma\omega_{\mathcal A}.
\]
The possible positive entry distances are therefore
\begin{equation}\label{eq:lars_entry}
 \gamma_{\rm entry}
 =
 \min_{\substack{n\notin\mathcal A\\\text{positive admissible values}}}
 \left\{
 \frac{C_{\rm corr}-c_n}{\omega_{\mathcal A}-v_n},
 \frac{C_{\rm corr}+c_n}{\omega_{\mathcal A}+v_n}
 \right\}.
\end{equation}
Ratios with zero denominators or nonpositive values are omitted. If there is no admissible ratio, set \(\gamma_{\rm entry}=+\infty\). The step is limited by the distance at which the active correlations vanish:
\begin{equation}\label{eq:lars_step}
 \gamma=\min\left\{
 \gamma_{\rm entry},\frac{C_{\rm corr}}{\omega_{\mathcal A}}
 \right\}.
\end{equation}

\paragraph{Step 3: update the fitted data and active set.}
Set
\begin{equation}\label{eq:lars_update}
 \mu\leftarrow\mu+\gamma u_{\mathcal A},
 \qquad
 \boldsymbol\rho\leftarrow y-\mu.
\end{equation}
Equivalently, the active regression coefficients are updated by \(\beta_j\leftarrow\beta_j+\gamma s_j(w_{\mathcal A})_j\). If a new candidate reaches the active correlation level before that level vanishes, admit it to \(\mathcal A\), record its index, and recompute the direction for the next segment. The implementation limits the number of LARS updates to \(B\) and stops earlier if the path terminates; the fitting stage uses the actual returned support size \(N_{\rm sel}\).

Equations \eqref{eq:lars_direction}--\eqref{eq:lars_update} describe the nondegenerate path. Numerical rank checks can exclude dependent columns and shorten the returned path. The Lasso modification that removes an active variable when its coefficient crosses zero is not used.

For the first active column, LARS moves along its signed temperature response only until another candidate becomes equally correlated with the residual. OMP instead completes the least-squares fit on its current selected support before choosing the next candidate. This difference can produce different candidate orders when neighboring thermal responses are strongly correlated.

As with OMP, only the order in which candidates enter the active set \((n_1,\ldots,n_{N_{\rm sel}})\) is passed to the physical fitting stage. The number of selected LARS variables is not interpreted as the physical source count.

\subsection{Continuous fitting and source-count selection}

The selected grid locations provide initial guesses for the physical source models. For a trial count \(k\), take the first \(k\) entries of the chosen OMP or LARS path:
\begin{equation}\label{eq:initial_locations}
 z_j^{(0)}=\xi_{n_j},\qquad j=1,\ldots,k.
\end{equation}
With these locations fixed, initialize the strengths by bounded nonnegative linear least squares,
\begin{equation}\label{eq:amplitude_initialization}
 a^{(0)}
 \in\operatorname*{arg\,min}_{0\leq a_j\leq a_{\max}}
 \left\|
 \sum_{j=1}^k a_j\g(z_j^{(0)})-y
 \right\|_2^2.
\end{equation}
Unnormalized responses retain the physical strength coordinates. A small positive floor is applied to starting strengths before the joint fit to avoid vanishing initial location derivatives; its value is specified in Section~\ref{sec:protocol}.

Starting from these parameters, solve
\begin{equation}\label{eq:fit}
 \min_{\substack{z_j\in\mathcal Z\\0\leq a_j\leq a_{\max}}}
 \left\|F_k(\theta)-y\right\|_2^2.
\end{equation}
Locations may move throughout \(\mathcal Z\). The grid supplies the starting locations; the final positions are continuous variables.

For the parameter ordering \(\theta=(z_1,a_1,\ldots,z_k,a_k)\), the Jacobian is
\begin{equation}\label{eq:physical_jacobian}
 J_k(\theta)=
 \left[
 a_1D\g(z_1),\ \g(z_1),\ \ldots,\
 a_kD\g(z_k),\ \g(z_k)
 \right].
\end{equation}
A bounded trust-region least-squares solver uses this Jacobian to update all locations and strengths jointly. The basic procedure performs one local fit for each tested source count. 

For independent Gaussian measurement errors with known standard deviation \(\sigma\), define
\begin{equation}\label{eq:discrepancy}
 \begin{aligned}
 T_M&=F_{\chi^2_M}^{-1}(1-\alpha_{\rm disc}),\qquad
 \epsilon=\sigma\sqrt{T_M},\\
 k_{\max}
 &=\min\left\{
 K_{\rm cap},N_{\rm sel},
 \left\lfloor\frac{M}{d+1}\right\rfloor
 \right\},
 \end{aligned}
\end{equation}
where \(\alpha_{\rm disc}\in(0,1)\) is the prescribed noise-tail probability, \(F_{\chi^2_M}\) is the cumulative distribution function of the central $\chi^2$ distribution with \(M\) degrees of freedom, and \(K_{\rm cap}\) is a prescribed search ceiling. The bound involving \(M\) excludes models with more parameters than observations. It is necessary for a full-column-rank Jacobian, but it does not ensure that such rank is attained.

The reconstruction is selected as follows:
\begin{enumerate}[label=(\roman*),leftmargin=*]
 \item If \(\norm y_2\leq\epsilon\), return the zero-source model.
 \item Otherwise, test \(k=1,\ldots,k_{\max}\) in increasing order. For each \(k\), perform the initialization and continuous fit described above.
 \item Return the first converged fit satisfying
 \[
  \RSS_k=
  \norm{F_k(\widehat\theta_k)-y}_2^2\leq\epsilon^2.
 \]
 \item If no tested fit satisfies this condition, report that no fit was accepted within the tested range.
\end{enumerate}
The candidate budget, source-count ceiling, and noise-tail probability are specified in Section~\ref{sec:protocol}.

The discrepancy threshold is calibrated to the norm of the raw noise vector; fitted degrees of freedom are not subtracted. It does not quantify the probability of selecting the correct physical source count. The returned count is the smallest accepted size among the attempted local fits. Since these fits may reach different local minima, their residuals need not decrease with \(k\).

The finite-source representation, nonnegative fitted strengths, and bounded parameter region impose structural constraints. Selecting the first attempted model consistent with the noise threshold limits model size without an explicit penalty term.

An off-grid source can be approximated using several nearby dictionary columns, and local fits can leave some strengths at zero. We therefore report the number of sources included in the accepted fit, including any retained zero-strength sources, and assess parameter accuracy separately. Proposition~\ref{prop:count} establishes a sufficient condition for correct source counting when the required fit is attained.

\paragraph{When the discrepancy rule identifies the physical count.}

For continuous \(\g\) on a compact admissible region and a finite amplitude bound, define the observation sets
\[
 \mathcal F_k=\left\{\sum_{j=1}^k a_j\g(z_j):
 z_j\in\mathcal Z,\ 0\leq a_j\leq a_{\max}\right\},
 \qquad \mathcal F_0=\{0\}.
\]
Continuity of \(\g\) makes these sets compact, and they are nested because a zero-strength source can be added. Write \(\mathcal R_k(v)=\operatorname{dist}(v,\mathcal F_k)\). Here distance is measured in the Euclidean observation norm. The statement below concerns global distances; the implemented local fits are considered afterward.

\begin{proposition}
\label{prop:count}
Let the true distinct-source configuration have \(K_*\geq1\) positive sources satisfying the fitting bounds, and put \(y=f_*+e+\boldsymbol e_{\rm mod}\), where \(f_*\in\mathcal F_{K_*}\), \(\norm e_2\leq\eta\), and \(\norm{\boldsymbol e_{\rm mod}}_2\leq b_{\rm mod}\). Here \(\boldsymbol e_{\rm mod}\) denotes deterministic forward-model discrepancy. If
\begin{equation}\label{eq:count_gap}
 \eta+b_{\rm mod}\leq\epsilon,
 \qquad \mathcal R_{K_*-1}(f_*)>\epsilon+\eta+b_{\rm mod},
\end{equation}
then the smallest \(k\) satisfying \(\mathcal R_k(y)\leq\epsilon\) is \(K_*\). If a converged attempted \(K_*\)-source fit attains residual norm at most \(\epsilon\), a sequential local-fit procedure with \(K_*\leq k_{\max}\) that tests every smaller size also returns \(K_*\) under these conditions.
\end{proposition}

\begin{proof}
Distance to a nonempty set is 1-Lipschitz, so
\[
 |\mathcal R_k(y)-\mathcal R_k(f_*)|
 \leq\norm{e+\boldsymbol e_{\rm mod}}_2
 \leq\eta+b_{\rm mod}.
\]
For every \(k<K_*\), adding zero-strength sources shows that
\(\mathcal F_k\subseteq\mathcal F_{K_*-1}\). Taking the
infimum over the smaller set therefore yields
\[
 \mathcal R_k(f_*)
 =\inf_{f\in\mathcal F_k}\norm{f_*-f}_2
 \geq
 \inf_{f\in\mathcal F_{K_*-1}}\norm{f_*-f}_2
 =\mathcal R_{K_*-1}(f_*).
\]
Combining this inequality with \eqref{eq:count_gap}, we obtain
\[
 \mathcal R_k(y)
 \geq\mathcal R_k(f_*)-\eta-b_{\rm mod}
 \geq\mathcal R_{K_*-1}(f_*)-\eta-b_{\rm mod}
 >\epsilon.
\]
On the other hand, \(f_*\in\mathcal F_{K_*}\) implies
\[
 \mathcal R_{K_*}(y)
 \leq\norm{y-f_*}_2
 \leq\eta+b_{\rm mod}
 \leq\epsilon.
\]
Hence \(K_*\) is the smallest model size whose global
distance to \(y\) does not exceed \(\epsilon\).

For the sequential local-fit procedure, any feasible fitted
parameter \(\widehat\theta_k\) with \(k<K_*\) satisfies
\[
 \norm{F_k(\widehat\theta_k)-y}_2
 \geq\mathcal R_k(y)>\epsilon.
\]
Thus no attempted fit with fewer than \(K_*\) sources can
be accepted. By assumption, \(K_*\leq k_{\max}\), and the
attempted \(K_*\)-source fit converges with residual norm
at most \(\epsilon\). Since the procedure tests source counts
in increasing order, it therefore returns \(K_*\).
\end{proof}

For an exact forward model, taking \(\eta=\epsilon\)
reduces \eqref{eq:count_gap} to
\[
 \mathcal R_{K_*-1}(f_*)>2\epsilon.
\]
Thus, correct source-count selection is ensured when the
noiseless response is sufficiently separated from every
model with fewer sources. In the presence of forward-model
discrepancy, both the acceptance radius and the required
separation must also accommodate \(b_{\rm mod}\).

Such separation cannot hold uniformly over all configurations
of distinct positive sources. An additional source at a fixed
admissible location changes the response by \(a\g(z)\), which
vanishes as its strength \(a\) tends to zero. Similarly,
Proposition~\ref{prop:merge} shows that two sources approaching
one another converge in response to a single source with
their combined strength, provided this strength satisfies
the fitting bound. Consequently, positivity and distinctness
alone do not ensure uniform source-count recovery in the
presence of noise.

\section{Source sensitivity and spatial resolution}\label{sec:resolution}

This section examines which source properties can be determined from the temperature observations. We first study single-source identification and localization sensitivity when strength is unknown. We then distinguish rejection of a single-source model from recovery of individual sources in a close pair, establish a conditional estimate for aggregate parameters, and derive a local noise-dependent reference for single-source rejection.


\subsection{Single-source identification and location sensitivity}
For the whole-space model, ratios of readings at two times eliminate the unknown source strength. Under suitable sensor geometry, the resulting source--sensor distances uniquely determine the source location.

\begin{proposition}
\label{prop:single}
For the whole-space response and \(0<t_1<t_2\), \(H_d(r,t_1)/H_d(r,t_2)\) is strictly decreasing for \(r>0\). Exact readings at these two times at \(d+1\) affinely independent sensors therefore determine the location and strength of a single positive source uniquely, provided it does not coincide with a sensor.
\end{proposition}
\begin{proof}
Set \(x=r^2/(4\kappa)>0\) and \(p=d/2-2\). The substitution \(v=1/s\) in \eqref{eq:whole} transforms the response ratio into
\[
 Q(x)=\frac{\int_{1/t_1}^\infty v^pe^{-xv}\,dv}{\int_{1/t_2}^\infty v^pe^{-xv}\,dv}.
\]
On \([1/t_2,\infty)\), normalize \(v^pe^{-xv}\) to a probability density and let \(V\) have that density. The threshold is \(\tau=1/t_1>1/t_2\) and \(Q(x)=\mathbb{P}_x(V\geq\tau)\). Differentiating the numerator and denominator yields \(Q'(x)=-\Cov_x(V,\boldsymbol1_{\{V\geq\tau\}})\). If \(p_+=\mathbb{P}_x(V\geq\tau)\), this covariance equals
\[
 p_+(1-p_+)\{\mathbb{E}_x[V\mid V\geq\tau]-\mathbb{E}_x[V\mid V<\tau]\}>0.
\]
Here \(\mathbb{P}_x\), \(\mathbb{E}_x\), and \(\Cov_x\) denote probability, expectation, and covariance under this normalized density. 
The density is positive on both sides of \(\tau\), and its exponential tail ensures finite moments for \(x>0\). Hence the covariance is strictly positive. Since \(x\) increases with \(r\), the response ratio is strictly decreasing in \(r\).

At sensor \(i\), the reading ratio cancels the strength and determines \(r_i=|b_i-z|\). Subtracting \(|b_1-z|^2=r_1^2\) from \(|b_i-z|^2=r_i^2\) yields
\[
 2(b_i-b_1)^Tz=|b_i|^2-|b_1|^2-r_i^2+r_1^2,\qquad i=2,\ldots,d+1.
\]
Affine independence ensures a unique solution for \(z\). Substitution into any positive reading then recovers the source strength.
\end{proof}

Proposition~\ref{prop:single} establishes single-source
identifiability from six temperature readings in two dimensions
and eight in three dimensions. The affine independence of
the sensors resolves the geometric ambiguity associated
with distance measurements. In contrast, if all sensors lie
in a hyperplane, reflection across that hyperplane leaves
every source--sensor distance unchanged. Two distinct
reflected source locations therefore produce identical
temperature histories whenever both are admissible,
regardless of the number of sampling times.

\paragraph{Local sensitivity with unknown strength.}

Where \(\g(z)\ne0\), define
\[
 q(z)=\g(z)/\norm{\g(z)}_2,\qquad
 \rho(z,z')=q(z)^Tq(z'),\qquad G(z)=Dq(z)^TDq(z).
\]
The matrix \(G\) is positive semidefinite and becomes a positive-definite local spatial metric only when \(Dq\) has full column rank.

\begin{proposition}
\label{prop:geometry}
If \(\g\) is \(C^3\) near \(z\) and nonzero there, then for a unit direction \(v\),
\begin{equation}\label{eq:rho}
 1-\rho(z,z+\delta v)=\frac{\delta^2}{2}v^TG(z)v+O(\delta^3),
 \qquad
 Dq=\frac{(I-qq^T)D\g}{\norm{\g}_2}.
\end{equation}
The remainder is uniform on compact source regions with bounded derivatives and \(\norm{\g}_2\) bounded away from zero.
\end{proposition}
\begin{proof}
Differentiating \(q^Tq=1\) once and twice in the direction \(v\)
establishes
\[
 q^TDq\,v=0,\qquad q^TD^2q[v,v]=-\norm{Dq\,v}_2^2.
\]
Taking the inner product of the Taylor expansion of
\(q(z+\delta v)\) with \(q(z)\) therefore leads to
\[
 \rho(z,z+\delta v)
 =1-\frac{\delta^2}{2}\norm{Dq(z)v}_2^2+O(\delta^3).
\]
This proves the first identity in \eqref{eq:rho}.
Differentiation of \(\g/\norm{\g}_2\) verifies the formula for \(Dq\).
On compact regions satisfying the stated bounds, the third
derivatives of \(q\) are uniformly bounded, so the remainder is uniform.
\end{proof}

The quantity \(v^TG(z)v\) measures the change in response shape caused by a displacement in direction \(v\). Absolute sensitivity also depends on source strength, response magnitude, and noise.

For the single-source Gaussian model, write $J_1=[aD\g(z),\g(z)]$
and let $\mathcal I=J_1^TJ_1/\sigma^2$ be the Fisher information, partitioned into position and strength blocks. Since $\g(z)\ne0$ makes the scalar strength block positive, define
\[
 \mathcal I_z=\mathcal I_{zz}
 -\mathcal I_{za}\mathcal I_{aa}^{-1}\mathcal I_{az}.
\]
This Schur complement measures the position information remaining when strength is free to adjust.

\begin{proposition}
\label{prop:profile}
Suppose that \(\g\) is continuously differentiable near an interior
parameter point \((z,a)\), with \(a>0\) and \(\g(z)\ne0\). Then
\begin{equation}\label{eq:profile}
 \mathcal I_z(z,a)
 =\frac{a^2}{\sigma^2}
   D\g(z)^T(I-q(z)q(z)^T)D\g(z)
 =\frac{a^2\norm{\g(z)}_2^2}{\sigma^2}G(z).
\end{equation}
For every \(v\in\R^d\),
\[
 v^T\mathcal I_z(z,a)v
 =\frac{1}{\sigma^2}
  \min_{s\in\R}
  \norm{aD\g(z)v+s\g(z)}_2^2.
\]
Moreover, \(\mathcal I_z(z,a)\) is positive definite if and only if
\(J_1(z,a)\) has full column rank.
\end{proposition}

\begin{proof}
The blocks of the joint Fisher information are
\[
 \mathcal I_{zz}=\frac{a^2}{\sigma^2}D\g^TD\g,\qquad
 \mathcal I_{za}=\frac{a}{\sigma^2}D\g^T\g,\qquad
 \mathcal I_{aa}=\frac{\norm{\g}_2^2}{\sigma^2}.
\]
Substitution into the Schur complement and use of the derivative
identity in \eqref{eq:rho} establish \eqref{eq:profile}.
The scalar minimization has solution
\(s_*=-a\g^TD\g\,v/\norm{\g}_2^2\), which verifies the variational
formula. This minimum vanishes for some \(v\ne0\) precisely when
\(aD\g\,v+s\g=0\) for a suitable \(s\).
Because \(\g\ne0\), any nontrivial null vector of \(J_1\)
must have a nonzero location component. The minimum is therefore
positive for every \(v\ne0\) if and only if \(J_1\) has full column rank.
\end{proof}

A strength change absorbs the component of a position-induced response change parallel to $\g$; the orthogonal component provides localization information. Small eigenvalues of $\mathcal I_z$ identify weak directions, regardless of candidate-grid spacing. When \(\mathcal I_z\) is nonsingular, a regular efficient local estimator in the small-noise regime has approximate position covariance $\mathcal I_z^{-1}$. Thus the standard deviation projected onto a unit direction $v$ is approximately $\sqrt{v^T\mathcal I_z^{-1}v}$. If motion is restricted to that line, the corresponding scale is instead $(v^T\mathcal I_zv)^{-1/2}$.

\subsection{Close-source approximation and aggregate recovery}

For a positive cluster, let \(A_c=\sum_j a_j\) and \(c=A_c^{-1}\sum_j a_jz_j\). The weighted center cancels the first spatial moment, making a single source \((c,A_c)\) a natural approximation.

\begin{proposition}
\label{prop:merge}
Suppose that \(\g\) is \(C^2\) on a convex neighborhood containing the cluster and separated from the sensors, and that \(\norm{D^2\g(z)[v,w]}_2\leq C_2|v||w|\) there for all \(v,w\in\R^d\). Then
\begin{equation}\label{eq:merge}
 \norm{\sum_ja_j\g(z_j)-A_c\g(c)}_2
 \leq\frac{C_2}{2}\sum_ja_j|z_j-c|^2.
\end{equation}
For a pair separated by \(\delta\), the right-hand side is \(C_2a_1a_2\delta^2/(2A_c)\).
\end{proposition}
\begin{proof}
Set \(v_j=z_j-c\). Taylor's formula with integral remainder reads
\[
 \g(z_j)=\g(c)+D\g(c)v_j+
 \int_0^1(1-s)D^2\g(c+sv_j)[v_j,v_j]\,ds.
\]
After multiplication by \(a_j\) and summation, the first-order
term cancels because \(\sum_j a_jv_j=0\).
The Hessian bound then yields \eqref{eq:merge}.
For a pair, the identities
\[
 z_1-c=\frac{a_2}{A_c}(z_1-z_2),\qquad
 z_2-c=\frac{a_1}{A_c}(z_2-z_1)
\]
reduce the weighted spread to
\(\sum_j a_j|z_j-c|^2=a_1a_2\delta^2/A_c\).
\end{proof}

For single-source fitting, write \(f=F_1\), so that
\(f(\theta)=A\g(z)\) for \(\theta=(z,A)\).
At any interior parameter with full-column-rank \(J_1(\theta)\), define
\[
 P^\perp(\theta)
 =I_M-J_1(\theta)\{J_1(\theta)^TJ_1(\theta)\}^{-1}J_1(\theta)^T.
\]
For the aggregate parameter, use the abbreviations
\[
 \theta_c=(c,A_c),\qquad J_c=J_1(\theta_c),\qquad
 P_c^\perp=P^\perp(\theta_c).
\]
If the merged parameters are allowed to adjust, their fit absorbs
the component of the cluster signal tangent to the one-source
response family. The orthogonal component determines the leading local departure from the single-source model.

For a pair, fix \(a_1,a_2>0\), a center \(c\), and a unit direction \(v\).
With \(A_c=a_1+a_2\), parameterize the separation by
\[
 z_1=c-\frac{a_2}{A_c}\delta v,\qquad
 z_2=c+\frac{a_1}{A_c}\delta v,
\]
so that \(c\) is the strength-weighted center and
\(|z_2-z_1|=\delta\) for \(\delta\geq0\). Define
\[
 y_\delta=a_1\g(z_1)+a_2\g(z_2),\qquad
 \boldsymbol d_2=\frac{a_1a_2}{2A_c}D^2\g(c)[v,v],\qquad
 C=\norm{P_c^\perp\boldsymbol d_2}_2,
\]
where \(C\) is used whenever the full-rank condition holds.

\begin{proposition}[Local single-source residual for a close pair]
\label{prop:curvature}
Suppose that \(\theta_c\) is interior to the one-source fitting
region, that \(\g\) is \(C^4\) near \(c\), and that \(J_c\) has
full column rank. For all sufficiently small \(\delta\geq0\),
there is a strict local one-source least-squares minimizer
\(\theta_\delta\) near \(\theta_c\), with \(\theta_\delta\to\theta_c\),
whose residual norm satisfies
\begin{equation}\label{eq:curvature}
 R_{1,\mathrm{loc}}(\delta)
 :=\norm{f(\theta_\delta)-y_\delta}_2
 =C\delta^2+O(\delta^3).
\end{equation}
For equal strengths, the remainder is \(O(\delta^4)\).
If \(C>0\), the leading residual along this family of local minimizers is quadratic.
\end{proposition}

\begin{proof}
Expansion about \(c\) and cancellation of the first spatial moment
produce
\[
 y_\delta=f(\theta_c)+\delta^2\boldsymbol d_2+O(\delta^3).
\]
The stationary equation
\(Df(\theta)^T\{f(\theta)-y_\delta\}=0\)
has derivative \(J_c^TJ_c\) with respect to \(\theta\) at
\((\theta_c,0)\). This matrix is positive definite.
The implicit function theorem defines a nearby stationary branch;
continuity of the least-squares Hessian ensures that these
stationary points are strict local minimizers. Expanding the
stationary equation yields
\[
 \theta_\delta=\theta_c+
 \delta^2(J_c^TJ_c)^{-1}J_c^T\boldsymbol d_2+O(\delta^3).
\]
Substitution into \(f\) then isolates the orthogonal residual:
\[
 f(\theta_\delta)-y_\delta
 =-\delta^2P_c^\perp\boldsymbol d_2+O(\delta^3).
\]
The reverse triangle inequality establishes \eqref{eq:curvature}.
For equal strengths, the odd terms in the data expansion cancel,
so its remainder is \(O(\delta^4)\).
Since \(\theta_\delta-\theta_c=O(\delta^2)\), the nonlinear
parameter terms also contribute only \(O(\delta^4)\).
The parameter and residual expansions therefore have remainders
of that order.
\end{proof}

The projected coefficient in \eqref{eq:curvature} determines the leading departure from the local single-source model. If it vanishes, higher-order terms control the residual. The expansion applies to the nearby minimizer; a remote single-source fit may have a different residual. Recovering the two individual sources requires additional sensitivity within the two-source model.

\begin{proposition}
\label{prop:cubic}
Let \(z_1=c_{\rm mid}-\delta v/2\), \(z_2=c_{\rm mid}+\delta v/2\), with
\(\delta>0\), \(|v|=1\), and \(a_1,a_2>0\).
Suppose that \(\g\) is \(C^3\) on a neighborhood of the segment
joining \(z_1\) and \(z_2\), and that
\[
 \sup_{|s|\leq\delta/2}
 \norm{D^3\g(c_{\rm mid}+sv)[v,v,v]}_2\leq C_3.
\]
For the two-source parameter
\(\theta=(z_1,a_1,z_2,a_2)\), use \(J_2(\theta)\) from \eqref{eq:physical_jacobian} and define
\[
 w_\delta=
 \left(\frac{\delta v}{2a_1},1,
       \frac{\delta v}{2a_2},-1\right).
\]
Then, writing \(J_2=J_2(\theta)\),
\begin{equation}\label{eq:cubic}
 \inf_{|w|=1}\norm{J_2w}_2
 \leq\frac{\norm{J_2w_\delta}_2}{|w_\delta|}
 \leq\frac{C_3\delta^3}{12|w_\delta|}
 \leq\frac{C_3\delta^3}{12\sqrt2}.
\end{equation}
\end{proposition}

\begin{proof}
Set \(\varphi(s)=\g(c_{\rm mid}+sv)\) and \(h=\delta/2\).
Multiplication by the Jacobian and two integrations by parts
establish the identity
\[
 J_2w_\delta
 =\varphi(-h)-\varphi(h)+h\{\varphi'(-h)+\varphi'(h)\}
 =\frac12\int_{-h}^h(h^2-s^2)\varphi^{(3)}(s)\,ds.
\]
Hence
\[
 \norm{J_2w_\delta}_2
 \leq\frac{C_3}{2}\int_{-h}^h(h^2-s^2)\,ds
 =\frac{2C_3h^3}{3}
 =\frac{C_3\delta^3}{12}.
\]
Testing the infimum with \(w_\delta/|w_\delta|\) and noting
that \(|w_\delta|\geq\sqrt2\) completes the proof.
\end{proof}

The geometric midpoint $c_{\rm mid}$ need not equal the strength-weighted center $c$. The direction $w_\delta$ transfers strength and shifts locations while preserving total strength and first spatial moment to first order. The linearized response change along this unit direction is bounded by a constant times the cube of separation, showing how individual-parameter sensitivity can remain weak even when the pair has a nonzero quadratic departure from a nearby single-source model. The construction supplies an upper bound on the smallest Jacobian gain, rather than a two-sided asymptotic rate. Related near-collision conditioning occurs in other super-resolution problems \cite{Batenkov2021}.

An accepted merged source may still provide useful estimates of
total strength and weighted center. Such recovery requires local
sensitivity of the one-source map in addition to a small
merged-response error.

\begin{proposition}
\label{prop:aggregate}
Let \(\theta_c=(c,A_c)\) be the aggregate parameter of a positive
cluster, and suppose that
\[
 s_c=\inf_{|w|=1}\norm{J_cw}_2>0.
\]
Suppose that \(f\) is continuously differentiable on an open
convex neighborhood \(U\) of \(\theta_c\), contained in the
interior of the admissible parameter region, and that
\[
 \sup_{\theta\in U}
 \norm{Df(\theta)-Df(\theta_c)}_{\rm op}
 \leq s_c/2.
\]
Let
\[
 y=\sum_j a_j\g(z_j)+e,\qquad \norm e_2\leq\eta,
\]
and assume the hypotheses of Proposition~\ref{prop:merge}.
If a fitted one-source parameter \(\widehat\theta\in U\)
satisfies
\(\norm{f(\widehat\theta)-y}_2\leq\epsilon\), then
\begin{equation}\label{eq:aggregate_stability}
 |\widehat\theta-\theta_c|
 \leq\frac{2}{s_c}
 \left(
  \epsilon+\eta
  +\frac{C_2}{2}\sum_j a_j|z_j-c|^2
 \right).
\end{equation}
\end{proposition}

\begin{proof}
Let \(h=\widehat\theta-\theta_c\). Convexity of \(U\) permits
integration along the segment from \(\theta_c\) to \(\widehat\theta\):
\[
 f(\widehat\theta)-f(\theta_c)
 =J_ch+\int_0^1\{Df(\theta_c+th)-J_c\}h\,dt.
\]
The definition of \(s_c\) and the derivative bound imply
\[
 \norm{f(\widehat\theta)-f(\theta_c)}_2
 \geq s_c|h|-\frac{s_c}{2}|h|=\frac{s_c}{2}|h|.
\]
The fitting criterion, the noise bound, and
Proposition~\ref{prop:merge} also imply
\[
 \norm{f(\widehat\theta)-f(\theta_c)}_2
 \leq\epsilon+\eta+\frac{C_2}{2}\sum_j a_j|z_j-c|^2.
\]
Combining these inequalities proves \eqref{eq:aggregate_stability}.
\end{proof}

The estimate separates the effects of fitting residual, measurement error, and cluster spread. It requires the fitted parameter to lie in the stated neighborhood of \(\theta_c\); a small residual alone does not ensure this condition.
The factor \(1/s_c\) quantifies error amplification when the
aggregate-source map is weakly sensitive to its parameters.

\subsection{A noise-dependent reference for rejecting one source}

For two specified mean vectors \(f_0,f_1\) under noise \(\mathcal N(0,\sigma^2I_M)\), the optimal equal-prior average testing error is
\[
 \Phi\!\left(-\frac{\norm{f_1-f_0}_2}{2\sigma}\right),
\]
where \(\Phi\) is the standard normal distribution function. Indeed, the likelihood-ratio test projects onto \(f_1-f_0\) and thresholds at the midpoint of the two projected means. A response difference of order \(\delta^2\) therefore provides a small Gaussian discrimination signal when its norm is small compared with \(\sigma\). This statement compares two specified configurations; it is not a minimax rate for the composite count problem.

Retain the pair parameterization, \(f\), \(\theta_c\), \(J_c\),
\(P_c^\perp\), \(\boldsymbol d_2\), and \(C\) from
Proposition~\ref{prop:curvature}, together with its assumptions.
The observation design, strengths, center, and direction remain fixed
as separation and noise tend to zero.
The local noiseless residual is \(C\delta^2+O(\delta^3)\), and
the discrepancy rule in \eqref{eq:discrepancy} uses the threshold
\(\sigma^2T_M\).
Locally, fitting removes the tangent component of the data
perturbation, leaving its projection onto the normal space.
The tangent approximation from nonlinear regression
\cite{BatesWatts1988} leads to the following local rejection limit.

\begin{proposition}[Local Gaussian rejection limit]
\label{prop:prob}
Let \(M>d+1\) and retain the assumptions and fixed quantities above. Suppose that \(\delta\to0\) and \(\sigma\to0\), with
\[
 \frac{\delta^2}{\sigma}\to\ell\in[0,\infty).
\]
For
\[
 y=a_1\g(z_1)+a_2\g(z_2)+\sigma Z,
 \qquad Z\sim\mathcal N(0,I_M),
\]
let \(\widehat\theta\) be the local one-source least-squares fit
near \(\theta_c\).
Writing \(\chi^2_r(\lambda)\) for a noncentral chi-square random
variable with \(r\) degrees of freedom and noncentrality
\(\lambda\), we have
\begin{equation}\label{eq:ncx}
 \frac{\norm{f(\widehat\theta)-y}_2^2}{\sigma^2}
 \ \Longrightarrow\
 \chi^2_{M-d-1}(\ell^2C^2).
\end{equation}
Here \(\Longrightarrow\) denotes convergence in distribution.
Consequently,
\[
 \mathbb{P}\!\left\{
 \norm{f(\widehat\theta)-y}_2^2>\sigma^2T_M
 \right\}
 \longrightarrow
 \mathbb{P}\!\left\{
 \chi^2_{M-d-1}(\ell^2C^2)>T_M
 \right\}.
\]
\end{proposition}

\begin{proof}
At $y_0=f(\theta_c)$, the stationary equation $Df(\theta)^T\{f(\theta)-y\}=0$ has derivative $J_c^TJ_c$ in $\theta$. The implicit function theorem defines a $C^2$ local minimizing map $\widehat\theta(y)$. Its residual $\boldsymbol r(y)=f(\widehat\theta(y))-y$ satisfies $\boldsymbol r(y_0)=0$ and $D\boldsymbol r(y_0)=-P_c^\perp$. Hence, with
\[
 \boldsymbol u_\delta=y-y_0
 =\delta^2\boldsymbol d_2+\sigma Z+O(\delta^3),
\]
we have $\boldsymbol r(y)=-P_c^\perp\boldsymbol u_\delta+O(\norm{\boldsymbol u_\delta}_2^2)$. The scaling assumption entails $\norm{\boldsymbol u_\delta}_2=O_{\mathbb{P}}(\sigma)$ and $\delta^3/\sigma\to0$, so
\[
 \frac{\boldsymbol r(y)}{\sigma}
 =-P_c^\perp\left(Z+\frac{\delta^2}{\sigma}\boldsymbol d_2\right)
 +o_{\mathbb{P}}(1)\ \Longrightarrow\ -P_c^\perp(Z+\ell\boldsymbol d_2).
\]
Here \(O_{\mathbb{P}}(\sigma)\) denotes boundedness in probability after division by \(\sigma\), and \(o_{\mathbb{P}}(1)\) denotes convergence to zero in probability. The local neighborhood is reached with probability tending to one, so any extension of the fit outside it leaves the limit unchanged. Since $P_c^\perp$ has rank $M-d-1$, the limiting squared norm is noncentral chi-square with noncentrality $\ell^2C^2$. The limiting distribution has no atom at $T_M$, which establishes convergence of the rejection probability.
\end{proof}

At finite noise, the limit suggests the approximation
\begin{equation}\label{eq:power}
 p_{\rm reject}\approx
 \mathbb{P}\!\left\{
 \chi^2_{M-d-1}(C^2\delta^4/\sigma^2)>T_M
 \right\}.
\end{equation}
Equation~\eqref{eq:power} concerns rejection of the local
one-source fit under the interior, full-rank, and exact-model
assumptions. It does not quantify the probability of selecting two sources or recovering their individual parameters.

\begin{proposition}
\label{prop:scale}
Retain the assumptions of Proposition~\ref{prop:prob} and
suppose that \(C>0\).
Choose
\[
 \pi\in
 \left(
 \mathbb{P}\{\chi^2_{M-d-1}(0)>T_M\},\,1
 \right).
\]
Let \(\lambda_\pi>0\) be the unique solution of
\[
 \mathbb{P}\{\chi^2_{M-d-1}(\lambda_\pi)>T_M\}=\pi.
\]
Define the local reference separation by
\begin{equation}\label{eq:scale}
 \delta_\pi(\sigma)
 =\left(\frac{\lambda_\pi\sigma^2}{C^2}\right)^{1/4}.
\end{equation}
For \(\delta=\delta_\pi(\sigma)\), the probability of rejecting
the local one-source fit tends to \(\pi\) as \(\sigma\to0\).
\end{proposition}

\begin{proof}
At a fixed positive threshold, the noncentral chi-square
upper-tail probability is continuous and strictly increasing
in the noncentrality parameter, from its central value to one.
Thus \(\lambda_\pi\) exists and is unique.
By definition,
\[
 \frac{\delta_\pi(\sigma)^2}{\sigma}
 =\frac{\sqrt{\lambda_\pi}}{C}.
\]
The asserted limit now follows from Proposition~\ref{prop:prob}.
\end{proof}

For fixed design and pair parameters with $C>0$, \eqref{eq:scale} scales as $\sqrt\sigma$: reducing the noise standard deviation fourfold halves the local reference separation. At fixed center, direction and total strength, a weaker companion reduces $C$ through $a_1a_2/A_c$.

The criteria address different tasks: $\mathcal I_z$ describes localization, $C$ describes the departure from one source, and $J_2$ describes individual pair parameters. Appending independent observations at unchanged noise variance cannot decrease position information after allowing strength to adjust or $C$; relocating sensors at a fixed reading count has no such monotonicity guarantee and is examined numerically below. Both monotonicity statements follow from least-squares variational representations: additional observations contribute nonnegative terms before minimization over the same parameter adjustment.

\section{Numerical assessment}\label{sec:numerics}

Four examples examine source recovery under different observation conditions. Example~1 studies separated sources, including the effects of sensor configuration, temporal sampling, noise, and initialization. Example~2 examines close pairs and aggregate recovery. Example~3 assesses the local single-source rejection approximation and compares sensor layouts. Example~4 considers a finite plate and three-dimensional reconstruction. Table~\ref{tab:protocol} summarizes the baseline configurations. All observations are synthetically generated.

\subsection{Common protocol and reporting criteria}\label{sec:protocol}

All variables are dimensionless and \(\kappa=1\). The source bounds are \(\mathcal Z=[-0.7,0.7]^d\) and \(0\leq a_j\leq8\). The candidate grid covers \([-0.65,0.65]^d\), with \(21^2\) nodes in two dimensions and \(13^3\) in three, with spacings \(h=0.065\) and \(1.3/12\), respectively. The initialization grid lies inside the admissible fitting region. Selection uses \(B=\min(10,M-1)\), with \(\tau_{\rm corr}=10^{-13}\) for OMP; the tested count is capped at \(\min(6,\lfloor M/(d+1)\rfloor,N_{\rm sel})\).   
The discrepancy threshold uses \(\alpha_{\rm disc}=0.01\), and starting strengths below \(10^{-7}\) are raised to that value before joint fitting.
All compared reconstruction methods use the same location and
strength bounds. OMP, LARS, and random initialization use the
source-count limit above. An additional benchmark is based on
alternating descent conditional gradient (ADCG)
\cite{Boyd2017}, which alternates conditional-gradient source
selection with local refinement of the current source configuration.
The ADCG-style benchmark uses the same absolute count ceiling of six;
its modifications are described below. Except for the explicitly identified diagnostics using the true parameters,
the reconstruction algorithms receive neither the true source count
nor the true source parameters.

A configuration denoted by $S\times T$ uses $S$ sensors and
$T$ sampling times per sensor, giving $M=ST$ scalar readings.
The two-dimensional sensors, in their selection order, are
\[
 b_1=(0,1),\quad
 b_2=(1/\sqrt2,-1/\sqrt2),\quad
 b_3=(-1/\sqrt2,-1/\sqrt2),\quad b_4=(0,-1).
\]
In three dimensions, the four sensors are
\[
 \frac{1.4}{\sqrt3}(1,1,1),\quad
 \frac{1.4}{\sqrt3}(1,-1,-1),\quad
 \frac{1.4}{\sqrt3}(-1,1,-1),\quad
 \frac{1.4}{\sqrt3}(-1,-1,1).
\]
They form a regular tetrahedron centered at the origin.
The 32 available times are logarithmically spaced on $[0.025,2]$;
the nested selection rule below determines the subset for each configuration.

For each source configuration, the noise standard deviation is set by
\begin{equation}\label{eq:noiseprotocol}
 \sigma=\nu\norm{y_{\rm ref}^{\rm clean}}_2/\sqrt{M_{\rm ref}},
\end{equation}
where $\nu$ is the prescribed relative noise level on the reference configuration and
$y_{\rm ref}^{\rm clean}$ contains the noiseless temperatures on a
fixed reference configuration: $3\times8$ in two dimensions and
$4\times8$ in three, with $M_{\rm ref}$ readings.
Within each thermal model, $\sigma$ is kept unchanged when comparing
observation configurations for the same sources and noise level.
Readings at shared sensor--time pairs use the same standardized noise,
providing paired comparisons. Unless stated otherwise, all reported noise percentages refer to the reference level $\nu$.

Each baseline setting contains 20 trials, paired across the compared methods and observation configurations. Separated sources have strengths uniform on \([0.7,1.3]\) and positions in a radius-\(0.6\) disk or ball with distances at least \(0.3\). Close pairs have unit strengths, a random center in \([-0.15,0.15]^2\), and a random direction. Center, direction, and noise are paired across separations. The minimum separation is used only to generate the separated-source test family.

A count success requires an accepted model size equal to the true count; retained zero-strength sources are included in that size. A joint success additionally requires a maximum matched position error \(E_z\leq h\) and a relative matched strength error \(E_a\leq0.2\), with matching that minimizes the total location distance. Specifically, for a correct-count fit let \(\varpi\) minimize \(\sum_j|\widehat z_{\varpi(j)}-z_j|\) over permutations, and define
\[
 E_z=\max_j|\widehat z_{\varpi(j)}-z_j|,\qquad
 E_a=\frac{\sum_j|\widehat a_{\varpi(j)}-a_j|}{\sum_j a_j}.
\]
All trials, including failures, remain in the success denominator. For accepted fits with positive total strength, put \(c=\sum_j a_jz_j/\sum_j a_j\) and \(\widehat c=\sum_j\widehat a_j\widehat z_j/\sum_j\widehat a_j\). The aggregate errors are
\[
E_A=\frac{|\sum_j\widehat a_j-\sum_j a_j|}{\sum_j a_j},\qquad E_c=|\widehat c-c|.
\]
Table~\ref{tab:close} also reports the first Wasserstein distance between the true and fitted normalized source distributions:
\[
 W_1(\mu_*,\widehat\mu)=\inf_{\gamma\in\Pi(\mu_*,\widehat\mu)}
 \int|z-z'|\,d\gamma(z,z'),\quad
 \mu_*=\frac{\sum_j a_j\delta_{z_j}}{\sum_j a_j},\quad
 \widehat\mu=\frac{\sum_j\widehat a_j\delta_{\widehat z_j}}{\sum_j\widehat a_j},
\]
where \(\Pi\) is the set of probability couplings with the specified marginals. Normalizing the strengths removes total-strength differences from this distance; these are measured separately by \(E_A\). The transport distance remains meaningful when the fitted and true source counts differ. Aggregate summaries include only accepted fits with positive total strength; the conditioning is stated with each result. The same success criteria are used throughout, with position tolerance given by the grid spacing in the corresponding dimension. In the success tables, an entry such as 18/17 denotes 18 count successes and 17 joint successes, both out of 20 trials; it is not a success ratio.

\paragraph{Fitting and time sampling.}

The bounded trust-region solver uses analytic Jacobians, Jacobian-based parameter scaling, at most 250 function evaluations for each tested source count, and function, step, and gradient tolerances \(10^{-10}\) after residual scaling by \(\max(\norm y_2,10^{-15})\). A fit must satisfy the solver convergence criterion as well as the discrepancy bound to be accepted. For example,
\[
\partial_rH_2=-\frac{\exp[-r^2/(4\kappa t)]}{2\pi\kappa r},\qquad
\partial_{z_\alpha}H_d=\partial_rH_d\frac{z_\alpha-b_\alpha}{r}.
\]
Plate derivatives include reflection signs. There is no physical boundary at the artificial source search box.

Let $\tau_0<\cdots<\tau_{31}$ be the 32 predefined sampling times. We first select $\tau_0$ and $\tau_{31}$. If $i<j$ are consecutive selected indices, choose the largest gap $j-i$ and add $\tau_{\lfloor(i+j)/2\rfloor}$; ties are resolved by choosing the earlier gap. Repeat until the required number of times is selected. For a fixed sensor set, each larger sampling set therefore contains the smaller sets. Configurations with different sensor counts but the same total number of readings use different sampling-time subsets. The observation-count and noise rows of Table~\ref{tab:protocol} are used in Example~1, the close-pair row in Example~2, and the last two rows in Example~4. Example~3 uses the separate sensor-selection and test data specified in Section~\ref{sec:newvalidation}.

\begin{table}[htbp]\centering\small
\caption{Baseline test configurations for Examples 1, 2, and 4; 20 paired trials per setting.}\label{tab:protocol}
\begin{tabularx}{\textwidth}{lXXX}\toprule
Suite & Observation designs & Source configurations & Reference noise \(\nu\) \\\midrule
Observation count & \(3\times4\), \(3\times8\), \(3\times16\), \(3\times32\), \(4\times12\), \(4\times24\) & \(K=1,\ldots,5\) & \(0.1\%,1\%\) \\
Noise & \(3\times8\), \(4\times12\) & \(K=1,2,3\) & \(0.01,0.1,0.5,1,2,5\%\) \\
Close pair & \(3\times8\), \(4\times12\) & \(\delta=0.02,0.04,0.08,\) \(0.12,0.2,0.3,0.45\) & \(0.1,1,5\%\) \\
Bounded plate & \(3\times8\), \(4\times12\) & \(K=1,2,3\) & \(0.1\%,1\%\) \\
Three dimensions & \(4\times8\), \(4\times16\) & \(K=1,2,3\) & \(0.1,1,5\%\) \\
\bottomrule
\end{tabularx}
\end{table}

\subsection{Example 1: separated-source recovery}\label{sec:observations}
We compare additional sensors with additional times at fixed total reading count. Appending independent readings at unchanged noise variance cannot reduce local information: in the variational formulas, every nuisance-parameter choice receives a nonnegative additional squared-residual term. This does not ensure monotone recovery rates because the threshold and local fits also change.

\begin{figure}[p]\centering\includegraphics[width=\textwidth]{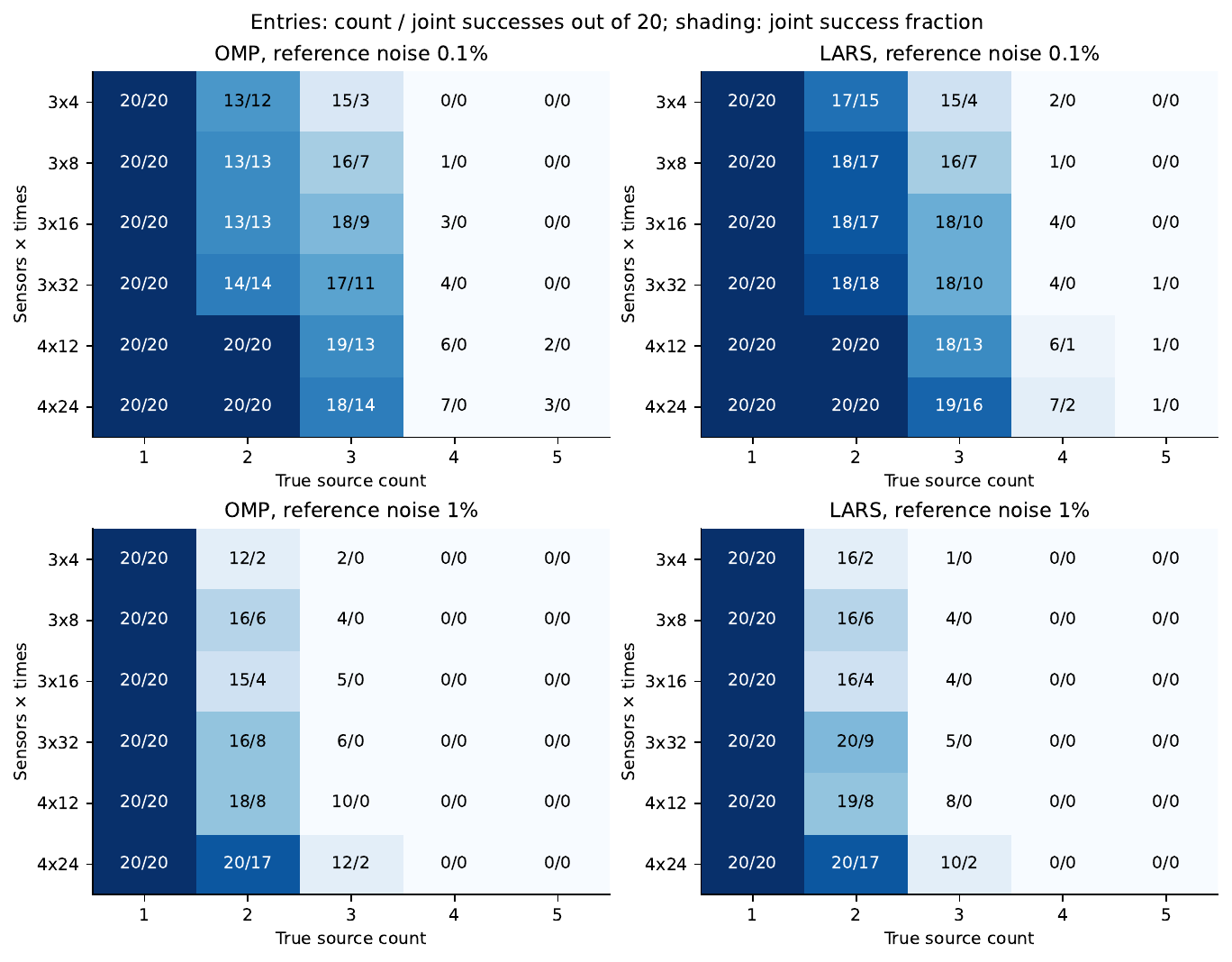}
\caption{Example 1: separated-source recovery across sensor and time-sampling configurations. Entries give count/joint successes out of 20; color shows the joint-success fraction.}\label{fig:observations}
\end{figure}

Figure~\ref{fig:observations} reports the observation tests in Example~1; the effect of noise is discussed below. Single sources are recovered in all baseline observation tests at \(0.1\%\) and \(1\%\) reference noise. With four sensors, two-source recovery is accurate at low noise; three-source recovery is more sensitive. For three sources and 48 readings at \(0.1\%\), LARS achieves count/joint \(18/10\) with \(3\times16\) and \(18/13\) with \(4\times12\). At 96 readings, the comparison is \(18/10\) versus \(19/16\). For these layouts, adding a fourth sensor improves joint recovery more than increasing the number of times at three sensors. The \(3\times4\) design cannot return five sources because its maximum tested source count is four.

\begin{figure}[htbp]\centering\includegraphics[width=\textwidth]{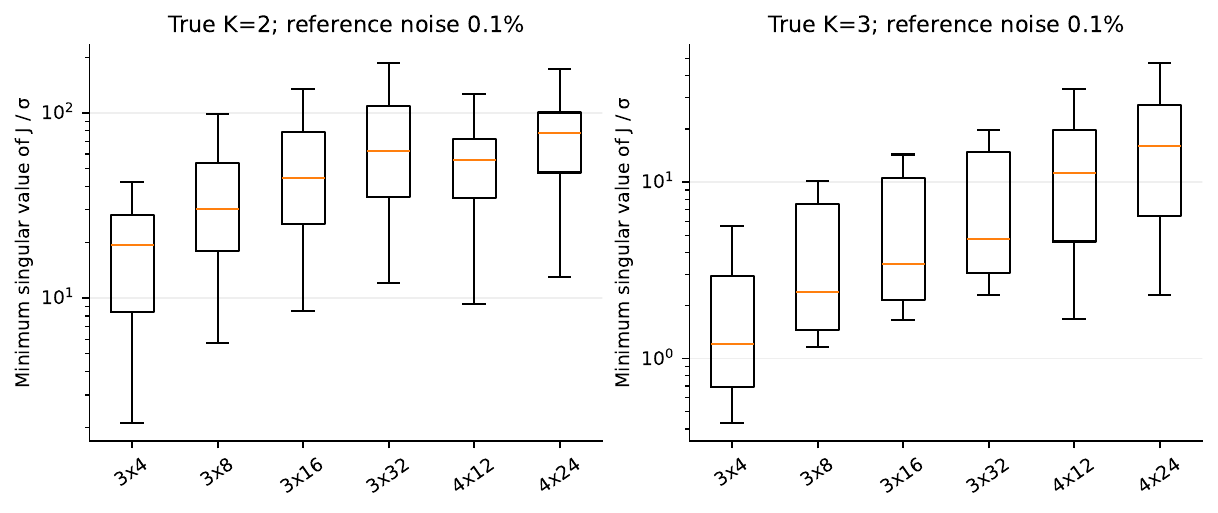}
\caption{Example 1: distributions of the smallest singular value of \(J_{K_*}(\theta_*)/\sigma\) across 20 separated-source configurations at \(0.1\%\) noise.}\label{fig:information}
\end{figure}

Figure~\ref{fig:information} reports the smallest singular values of the noise-scaled Jacobians for the same separated-source configurations as Figure~\ref{fig:observations}:
for three sources, the four-sensor designs have larger median values than three-sensor designs with the same total number of readings.
These values use the stated dimensionless parameter coordinates and depend on that scaling. They describe local sensitivity at the true parameters, rather than initialization, convergence, or global count identifiability.

As noise increases, individual-source accuracy deteriorates even
when temperature fits satisfy the discrepancy threshold.
With \(4\times12\) readings and three sources, LARS achieves 13/20
joint successes at \(0.1\%\) reference noise and none at \(1\%\);
all 20 higher-noise data vectors have accepted fits, but only eight
have the correct count. At \(0.01\%\), 16 fits are accepted and
14 meet the joint criterion. In the four unaccepted trials, the
true configuration satisfies the discrepancy bound:
three failures arise from the convergence requirement despite
below-threshold residuals in larger attempted models, and one from
above-threshold residuals in all attempted fits. Thus a tighter
noise threshold can expose limitations of local fitting and
termination even when a physical model satisfying the discrepancy bound exists.

\paragraph{Matched initialization and optional postprocessing.}

Random initialization uses ten randomly ordered grid candidates followed by the same strength initialization and continuous fits for successive source counts; the zero-source test is omitted in this positive-signal comparison. The ADCG-style benchmark uses grid-plus-local-search insertion and sequential least squares programming (SLSQP) correction \cite{Kraft1988} of all current locations and nonnegative strengths, with at most ten insertions and six retained sources. It removes strengths below $10^{-6}$ of the total. This is a bounded approximate variant of the ADCG principle, rather than an exact solution of its original measure-constrained problem. The basic selection-and-fitting procedure instead retains zero-strength sources, and work counts are not directly comparable across these methods. Fits initialized at the true parameters use the true count and parameters only to diagnose initialization effects.

Variable projection eliminates strengths by bounded nonnegative least squares and refines locations using a fixed-active-set residual Jacobian \cite{Golub1973}, with at most 150 evaluations. It starts from an accepted fit, preserves count, and is retained only if it converges and increases squared residual by no more than a relative $10^{-10}$. The implementation does not apply a separate discrepancy recheck after this update. Constraint changes can make the reduced objective nonsmooth.

Greedy merge/refit replaces a candidate pair by its summed strength at its weighted center, chooses the pair that minimizes the prediction change, and continuously refits the reduced model. Zero summed strength or a sum above $a_{\max}$ excludes a pair. A reduction requires convergence and the original discrepancy bound; otherwise the procedure stops without testing other pairs. No merging distance is imposed. The two postprocessors start separately from the same basic fit.

Table~\ref{tab:comparison} reports the matched reconstruction results and the diagnostic using the true parameters.

\begin{table}[htbp]\centering\small\caption{Example 1: initialization comparisons. Entries give count/joint successes out of 20; fits initialized at the true parameters report joint successes only.}\label{tab:comparison}\begin{tabular}{llrrrrr}\toprule Design & $K$/noise (\%) & OMP & LARS & Random & ADCG-style & True-start joint \\\midrule
$3\times8$ & 2/0.1 & 13/13 & 18/17 & 17/17 & 18/17 & 19/20 \\
$3\times8$ & 2/1 & 16/6 & 16/6 & 15/4 & 15/5 & 6/20 \\
$3\times8$ & 3/0.1 & 16/7 & 16/7 & 16/5 & 15/3 & 7/20 \\
$3\times8$ & 3/1 & 4/0 & 4/0 & 5/0 & 5/0 & 0/20 \\
$4\times12$ & 2/0.1 & 20/20 & 20/20 & 20/20 & 20/20 & 20/20 \\
$4\times12$ & 2/1 & 18/8 & 19/8 & 19/8 & 18/8 & 8/20 \\
$4\times12$ & 3/0.1 & 19/13 & 18/13 & 19/14 & 15/9 & 18/20 \\
$4\times12$ & 3/1 & 10/0 & 8/0 & 8/0 & 9/0 & 0/20 \\
\bottomrule\end{tabular}\end{table}
OMP, LARS, and random initialization perform similarly in most cells. The results do not show a consistent advantage of either sparse selector over random initialization with the same continuous-fitting protocol. For three sources with \(4\times12\) readings at \(0.1\%\) noise, fitting initialized at the true parameters increases joint successes from 13/20 for LARS to 18/20. At \(1\%\), neither design yields a joint success even from the true parameters. Favorable initialization therefore does not remove the loss of parameter accuracy at this noise level. Four and five sources are substantially harder: with \(4\times24\) readings at \(0.1\%\), LARS achieves count/joint successes of \(7/2\) and \(1/0\), respectively. Figure~\ref{fig:observations} summarizes the complete observation results.

Table~\ref{tab:post} pools the eight comparison settings in Table~\ref{tab:comparison}, giving 160 data vectors per selector. Variable projection changes neither the count nor the joint-success totals. Merge/refit increases joint successes from 67 to 73 for OMP and from 71 to 73 for LARS, with more function evaluations in both cases. The work totals include physical and reduced least-squares fits but exclude dictionary selection and linear strength solves.

\begin{table}[htbp]\centering\small\caption{Example 1: postprocessing results for 160 data vectors per selector. Work is the median number of least-squares function evaluations.}\label{tab:post}\begin{tabular}{llrrrr}\toprule Selector & Procedure & Accepted & Count & Joint & Work \\\midrule
OMP & Basic & 159 & 116 & 67 & 45 \\
OMP & Variable projection & 159 & 116 & 67 & 47 \\
OMP & Merge/refit & 159 & 123 & 73 & 51 \\
LARS & Basic & 159 & 119 & 71 & 35 \\
LARS & Variable projection & 159 & 119 & 71 & 37 \\
LARS & Merge/refit & 159 & 122 & 73 & 41.5 \\
\bottomrule\end{tabular}\end{table}

\FloatBarrier
\subsection{Example 2: close pairs and aggregate recovery}\label{sec:close_example}

\begin{figure}[htbp]\centering\includegraphics[width=\textwidth]{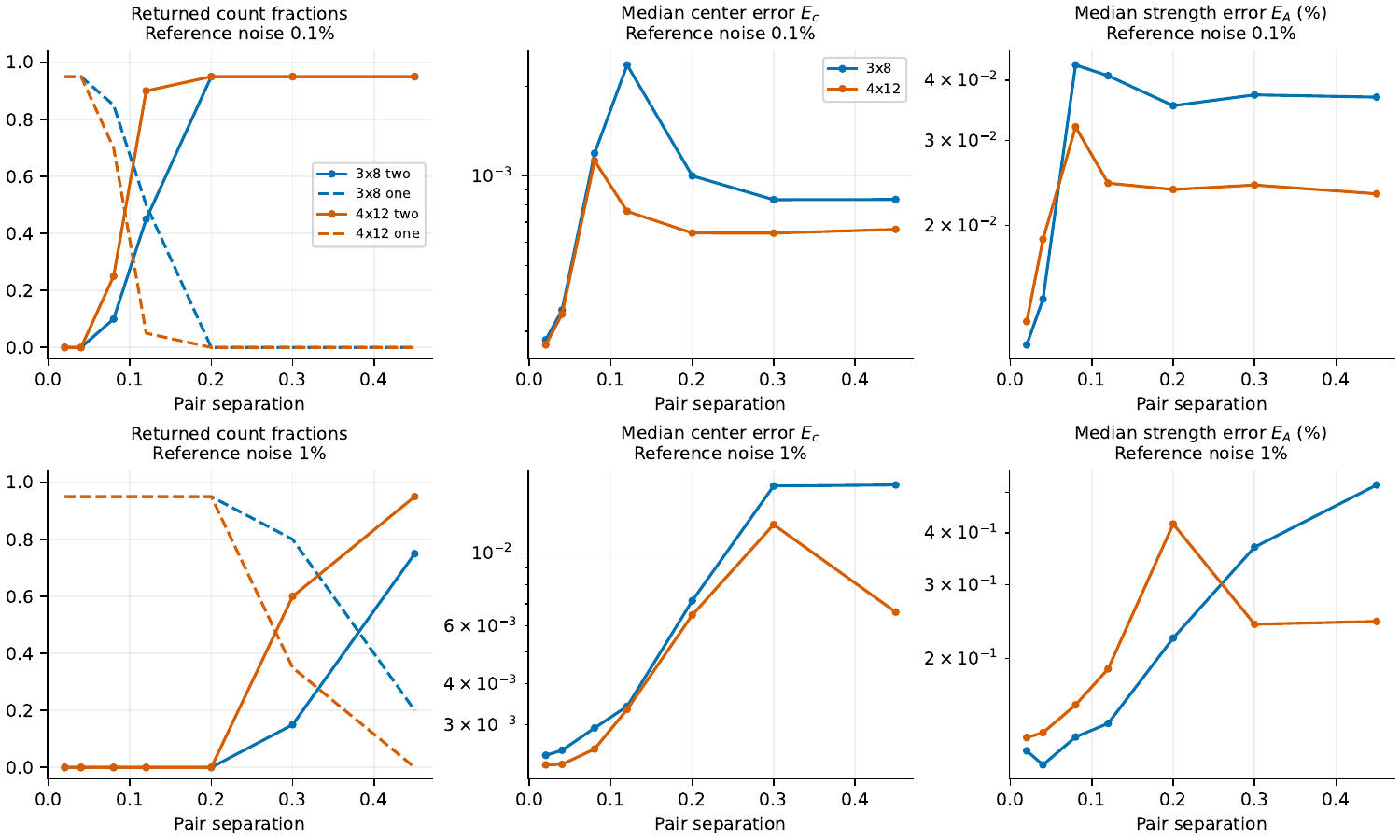}
\caption{Example 2: LARS close-pair recovery. Columns show the fractions returning one or two sources, median weighted-center errors, and median total-strength errors; rows correspond to \(0.1\%\) and \(1\%\) reference noise. Count fractions use all 20 trials; error medians are conditional on accepted fits with positive total strength.}\label{fig:close}
\end{figure}

Figure~\ref{fig:close} shows the close-pair tests at \(0.1\%\) and \(1\%\) noise. The count panels use all 20 trials; the aggregate-error panels use accepted fits only. At separation \(0.12\) and \(0.1\%\) reference noise, LARS returns two sources in 9/20 trials with \(3\times8\) and 18/20 with \(4\times12\), but joint successes are only 1/20 and 5/20. The gap between correct-count selection and accurate individual parameters is consistent with the distinction in Propositions~\ref{prop:curvature} and~\ref{prop:cubic}. At separation \(0.04\) and \(1\%\), each design returns an accepted merged source in 19/20 trials, with the aggregate errors in Table~\ref{tab:close}.

\begin{figure}[htbp]\centering
\includegraphics[width=\textwidth]{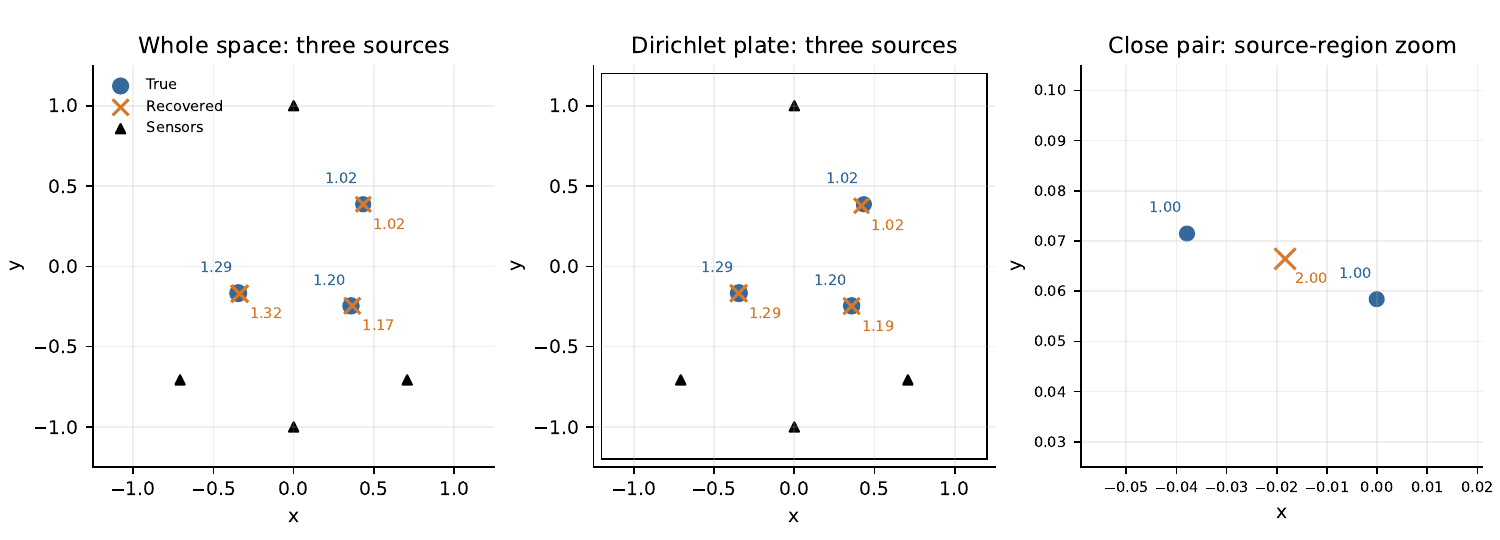}
\caption{Representative LARS reconstructions from Examples 1, 4, and 2: separated whole-space sources, finite-plate sources, and a merged close pair. Labels indicate source strengths.}\label{fig:recon}
\end{figure}

\begin{figure}[htbp]\centering\includegraphics[width=\textwidth]{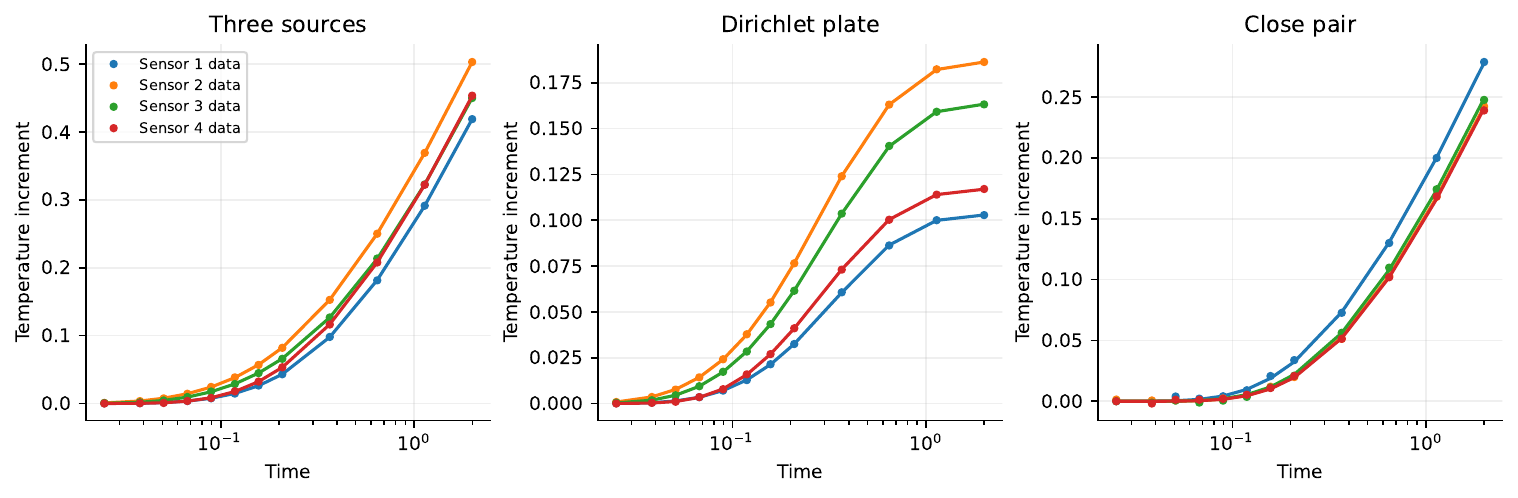}
\caption{Temperature observations (symbols) and LARS predictions (lines) for the three cases in Figure~\ref{fig:recon}. Colors distinguish sensors.}\label{fig:temperature}
\end{figure}

\begin{table}[htbp]\centering\small\caption{Example 2: merged-pair recovery at \(\delta=0.04\) and \(1\%\) noise. Errors are medians over 19 accepted one-source fits per design.}\label{tab:close}\begin{tabular}{lrrrr}\toprule Design & Merged / all & $E_c$ & $E_A$ (\%) & $W_1$ \\\midrule
$3\times8$ & 19/20 & 0.0025 & 0.111 & 0.0200 \\
$4\times12$ & 19/20 & 0.0023 & 0.133 & 0.0200 \\
\bottomrule\end{tabular}\end{table}
For an equal-strength pair, transport to a single source at the exact center gives \(W_1=\delta/2\). The last column of Table~\ref{tab:close} records \(W_1=0.0200\) for both designs, equal to this reference value at the reported precision. Thus the small center and total-strength errors describe accurate aggregate recovery, while the transport distance records the unresolved spatial spread. Figure~\ref{fig:recon} brings together one separated-source realization from Example~1, its finite-plate reconstruction from Example~4, and a merged-pair realization from Example~2. Both three-source cases use \(4\times12\) readings and \(0.1\%\) noise; the close pair uses \(\delta=0.04\) and \(1\%\) noise. Its spatial panel is magnified and omits the distant sensors. Table~\ref{tab:coordinates} lists the three-source positions and strengths after matching that minimizes total location distance; these estimates lie off the candidate grid. Figure~\ref{fig:temperature} compares their temperature predictions with the observations. Close agreement in temperature occurs for both the resolved three-source examples and the merged pair.

\begin{table}[htbp]\centering\scriptsize\caption{Examples 1 and 4: true and fitted parameters for the three-source cases in Figure~\ref{fig:recon}.}\label{tab:coordinates}\begin{tabular}{rrrrrrr}\toprule & \multicolumn{2}{c}{True} & \multicolumn{2}{c}{Whole-space LARS} & \multicolumn{2}{c}{Plate LARS} \\ Source & $(x,y)$ & $a$ & $(\widehat x,\widehat y)$ & $\widehat a$ & $(\widehat x,\widehat y)$ & $\widehat a$ \\\midrule
1 & $(-0.3452,-0.1667)$ & 1.2947 & $(-0.3362,-0.1703)$ & 1.3189 & $(-0.3467,-0.1673)$ & 1.2885 \\
2 & $(0.4329,0.3868)$ & 1.0183 & $(0.4337,0.3864)$ & 1.0216 & $(0.4184,0.3773)$ & 1.0154 \\
3 & $(0.3569,-0.2458)$ & 1.1971 & $(0.3652,-0.2456)$ & 1.1681 & $(0.3568,-0.2477)$ & 1.1897 \\
\bottomrule\end{tabular}\end{table}

\begin{figure}[htbp]\centering
\includegraphics[width=\textwidth]{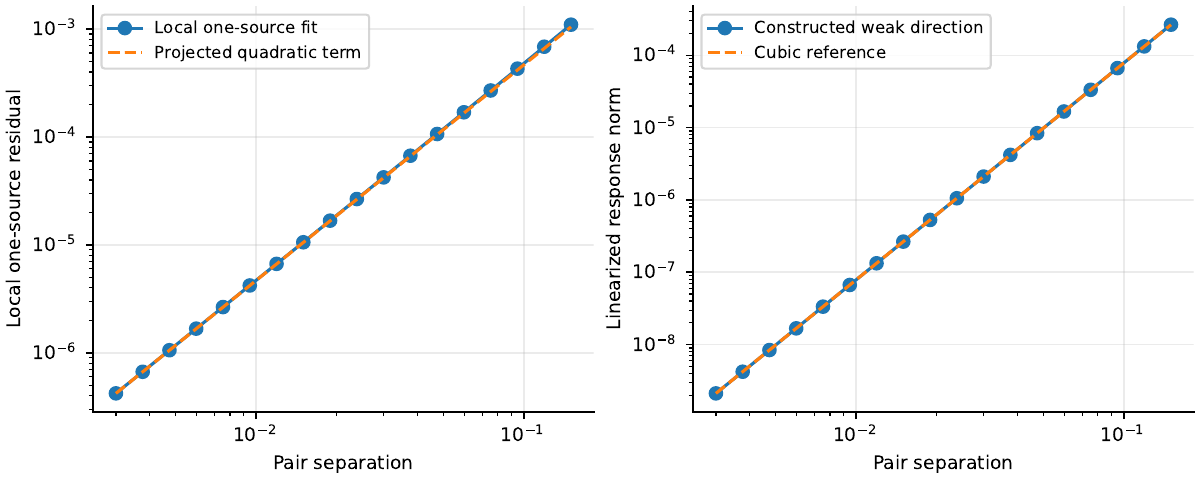}
\caption{Example 2: noiseless one-source residual and linearized response norm along the unit parameter direction \(w_\delta/|w_\delta|\).}\label{fig:resolution_scaling}
\end{figure}

The noiseless check in Figure~\ref{fig:resolution_scaling} uses four sensors, twelve nested times, $v=(\cos0.7,\sin0.7)$, strengths $(0.8,1.2)$ and separations $0.003$--$0.15$. The quadratic-residual test fixes the weighted center at $(0.08,-0.06)$, follows Proposition~\ref{prop:curvature}, and initializes one-source fits at the exact aggregate. The weak-direction test fixes the geometric midpoint at the same coordinates and uses positions $c_{\rm mid}\pm\delta v/2$. These are different configurations for unequal strengths. The projected quadratic coefficient is $C=0.0466748$; log--log slopes over the eight smallest separations are $2.0021$ for the local residual and $3.0000$ for the linearized response change along the constructed unit parameter direction. They support the quadratic expansion and demonstrate cubic behavior along the tested direction, without establishing a two-sided asymptotic estimate of the smallest Jacobian singular value.

\FloatBarrier
\subsection{Example 3: local rejection and sensor layout}\label{sec:newvalidation}

Two independent batches compare the local rejection approximation with reconstruction and sensor layout. Both use the two-dimensional whole-space model and retain the first three sensors from Section~\ref{sec:protocol}. Each layout has four sensors and twelve common sampling times, giving 48 scalar readings per data vector. For these designs, $T_M=73.6826$ and $\lambda_{0.9}=49.8359$ in \eqref{eq:scale}. Only the fourth sensor moves on the unit circle, on a $15^\circ$ grid excluding coincident positions; angles are counterclockwise from the positive first coordinate axis.

The first design search uses a $5\times5$ center grid on $[-0.4,0.4]^2$ and 36 directions. Maximizing the worst projected pair signal selects $30^\circ$; maximizing the smallest single-source position information after allowing strength to adjust selects $135^\circ$. Their training objectives improve by $22.7\%$ and $90.4\%$ over the default $270^\circ$ layout.
For a layout \(L\), let \(C_L(c,v)\) be the coefficient in
Proposition~\ref{prop:curvature} for strengths \((1,1)\), and
let \(q_L=\g_L/\norm{\g_L}_2\). The two objectives are
\[
 \max_L\min_{c,v}C_L(c,v),\qquad
 \max_L\min_c\lambda_{\min}\!\left(
 4D\g_L(c)^T(I-q_L(c)q_L(c)^T)D\g_L(c)\right).
\]
The minima are over the stated finite training sets.
The second matrix is the strength-two profiled location
information with the common noise factor omitted.
 The first test batch contains 16 groups with independently generated source centers and directions, centers uniform on $[-0.35,0.35]^2$, and angles uniform on $[0,\pi)$. Each group combines strength ratios $1:1$ and $1:3$ with total strength two, separations $0.06,0.12,0.2,0.36$, noise levels $0.1,0.5,1\%$, and six noise realizations, giving 2304 data vectors per layout.

The second design search uses a separate training set of 64 center/direction pairs and selects $30^\circ$ by mean predicted rejection in \eqref{eq:power} and $45^\circ$ by the mean logarithm of the smallest two-source Jacobian singular value normalized by the reference root-mean-square (RMS) response. The rejection objective averages \eqref{eq:power} over the
64 training center/direction pairs, strength pairs \((1,1)\)
and \((0.5,1.5)\), separations \(0.06,0.12,0.2,0.36\), and
reference noise levels \(0.1,0.5,1\%\), with equal weights.
The two-source objective averages
\(\log\max\{s_{\min}(J_{2,L})/r_{\rm ref},10^{-30}\}\)
over the same training pairs and strengths, using separations
\(0.12,0.2,0.36\), where
\(r_{\rm ref}=\norm{2\g_{\rm ref}(c)}_2/\sqrt{24}\).
Here \(s_{\min}(J)\) denotes the smallest singular value of \(J\).
The singular values use the parameter coordinates stated in
Section~\ref{sec:method}.
The selected layouts are fixed before testing on 32 new independent groups with four noise realizations per setting, giving 3072 data vectors per layout. In both batches, noise is calibrated on the merged strength-two response at the same center in the fixed $3\times8$ reference layout and held fixed across layouts, separations and strength ratios. Standardized noise is shared within each group. Thus the center/direction groups, rather than the individual data vectors, are statistically independent.

Reconstruction uses basic LARS and the common success criteria. Noiseless residual checks take the smallest converged one-source fit from three starts: the true weighted center and the two true source positions. Noisy single-source rejection uses the basic LARS fit; all such fits converged in these batches. These comparisons assess the approximation for the implemented fits, without certifying global optimality or membership in the local minimizing branch of Proposition~\ref{prop:curvature}.

The mean squared probability scores,
$n^{-1}\sum_{i=1}^n(p_i-o_i)^2$, are $0.0405$ and $0.0397$
in the two batches. The corresponding scores are $0.1060$ and
$0.1069$ for the deterministic prediction obtained by comparing the
noiseless residual with the discrepancy threshold.
Here $n$ is the number of evaluated data vectors, $p_i$ is the
predicted rejection probability, and $o_i=1$ denotes rejection
while $o_i=0$ denotes acceptance. The binary reference assigns
probability one or zero according to the noiseless-residual comparison.

\begin{figure}[htbp]\centering
\includegraphics[width=\textwidth]{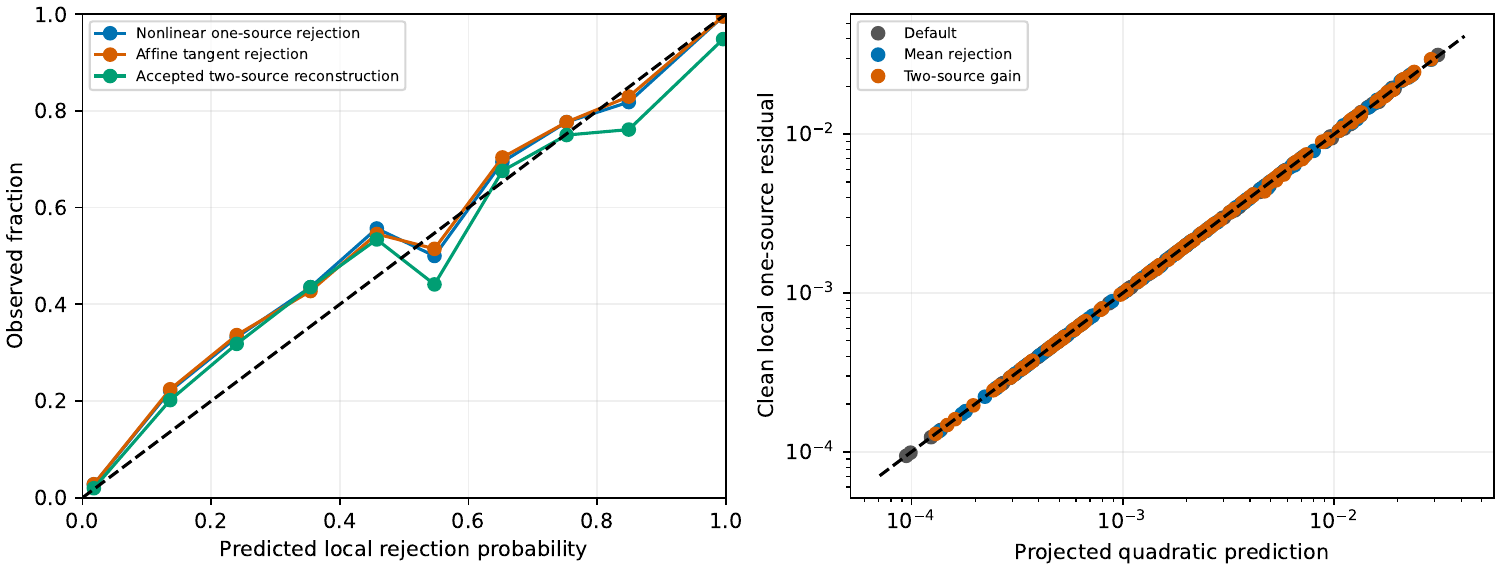}
\caption{Example 3, second test batch. Left: predicted single-source rejection probability versus binned rejection fractions for nonlinear and affine fits, and the fraction of accepted two-source reconstructions. Right: quadratic residual prediction and fitted noiseless residual. Dashed lines indicate equality.}\label{fig:calibration}
\end{figure}

In the second batch, mean predicted and observed rejection fractions are $0.4605$ and $0.4723$. Figure~\ref{fig:calibration} shows binned frequencies, the linearized one-source approximation, and accepted two-source counts. The latter do not require accurate individual parameters. Bins pool paired data vectors within these groups, and some depart from the identity line. The right panel compares the quadratic prediction with the diagnostic noiseless residual.

Tables~\ref{tab:design_first} and~\ref{tab:design_second} show that larger sensitivity objectives do not improve joint recovery in these tests. In the second batch, the mean-rejection design rejects one source in 1478 cases versus 1408 for the default layout, but does not increase joint recovery. Rejecting a single-source fit is therefore insufficient to ensure accurate two-source reconstruction.

\begin{table}[htbp]\centering
\caption{Example 3, first test batch: recovery totals for 2304 data vectors per layout, organized into 16 groups with independently generated source centers and directions.}\label{tab:design_first}
\begin{tabular}{lrrr}\toprule
Fourth sensor & Accepted & Two-source count & Joint parameters \\\midrule
Default \(270^\circ\) & 2304 & 1023 & 488 \\
Pair-sensitivity design \(30^\circ\) & 2290 & 961 & 386 \\
Single-source design \(135^\circ\) & 2293 & 848 & 405 \\\bottomrule
\end{tabular}\end{table}
In the first batch, neither designed layout improves count or joint recovery; the pair-design joint-rate difference is $-4.43$ percentage points. The independent sampling units are the 16 center/direction groups.
\begin{table}[htbp]\centering
\caption{Example 3, second test batch: recovery totals for 3072 data vectors per layout, organized into 32 groups with independently generated source centers and directions.}\label{tab:design_second}
\begin{tabular}{lrrr}\toprule
Fourth sensor & Accepted & Two-source count & Joint parameters \\\midrule
Default \(270^\circ\) & 3047 & 1377 & 636 \\
Mean-rejection design \(30^\circ\) & 3026 & 1385 & 616 \\
Two-source sensitivity design \(45^\circ\) & 3032 & 1355 & 596 \\\bottomrule
\end{tabular}\end{table}
In the second batch, mean-rejection design adds $8/3072$ correct counts ($0.26$ percentage points) but loses $20/3072$ joint successes. Neither design improves joint recovery; the comparison is descriptive, with 32 independent center/direction groups.

Across the three layouts, the second batch contains 3072 data vectors
at \(0.1\%\) noise, or 1024 per layout. For these data, fits initialized at the true two-source parameters give \(430\), \(438\), and \(441\) joint successes out of 1024 per layout. The corresponding basic-method joint-success counts are \(411\), \(413\), and \(399\). As in Table~\ref{tab:comparison}, the fits initialized at the true parameters measure the effect of a favorable initialization with known count.

\FloatBarrier

\subsection{Example 4: finite-plate and three-dimensional reconstruction}
\label{sec:transfer}

We apply the reconstruction procedure to a finite plate
$\mathcal D=(-1.2,1.2)^2$ with homogeneous Dirichlet boundary
conditions, and to the three-dimensional whole-space model.
For the plate, the method of images represents the response as
\begin{equation}\label{eq:images}
 R_{\mathcal D}(b,z,t)
 =\sum_{n\in\mathbb Z^2}
  \sum_{\varepsilon\in\{-1,1\}^2}
  \varepsilon_1\varepsilon_2
  H_2(|b-z_{n,\varepsilon}|,t),
\end{equation}
where
\[
 z_{n,\varepsilon}
 =\varepsilon\odot(z+1.2\boldsymbol1)
  +4.8n-1.2\boldsymbol1,
 \qquad \boldsymbol1=(1,1)^T.
\]
Here $\odot$ denotes componentwise multiplication,
$n$ indexes image translations, and $\varepsilon$ specifies
coordinate reflections. The inverse response retains
$n_1,n_2=-2,\ldots,2$. Sensors are placed strictly inside the plate because the homogeneous Dirichlet condition fixes the boundary temperature at zero, providing no source-dependent temperature observations there.

Temperature data are generated independently using a five-point
finite-difference discretization with $257^2$ interior nodes
and spacing $h_{\rm FD}=2.4/258$. Each off-grid source is
distributed among its four neighboring nodes using bilinear
weights scaled by $h_{\rm FD}^{-2}$, and sensor temperatures
are obtained by bilinear interpolation. Constant forcing is
integrated exactly for the spatially discretized system using
discrete sine modes. Thus, data generation and reconstruction
use different forward implementations.

\begin{table}[htbp]\centering\small\caption{Example 4: finite-plate and three-dimensional recovery. Entries give count/joint successes out of 20.}\label{tab:domain}\begin{tabular}{llrrrr}\toprule Model & Design & $K$ & Noise (\%) & OMP & LARS \\\midrule
Plate & $3\times8$ & 2 & 0.1 & 18/18 & 18/18 \\
Plate & $3\times8$ & 2 & 1 & 18/3 & 18/4 \\
Plate & $3\times8$ & 3 & 0.1 & 15/6 & 16/3 \\
Plate & $3\times8$ & 3 & 1 & 6/0 & 6/0 \\
Plate & $4\times12$ & 2 & 0.1 & 19/19 & 20/20 \\
Plate & $4\times12$ & 2 & 1 & 19/8 & 20/8 \\
Plate & $4\times12$ & 3 & 0.1 & 19/13 & 18/14 \\
Plate & $4\times12$ & 3 & 1 & 10/0 & 10/0 \\
3D & $4\times8$ & 2 & 0.1 & 15/10 & 18/12 \\
3D & $4\times8$ & 2 & 1 & 10/2 & 12/4 \\
3D & $4\times8$ & 3 & 0.1 & 13/3 & 17/3 \\
3D & $4\times8$ & 3 & 1 & 5/0 & 5/0 \\
3D & $4\times16$ & 2 & 0.1 & 15/10 & 18/12 \\
3D & $4\times16$ & 2 & 1 & 13/4 & 14/4 \\
3D & $4\times16$ & 3 & 0.1 & 13/2 & 16/2 \\
3D & $4\times16$ & 3 & 1 & 4/0 & 4/0 \\
\bottomrule\end{tabular}\end{table}

Table~\ref{tab:domain} reports count and joint recovery at
$0.1\%$ and $1\%$ noise. Single-source cases achieve 20/20
successes for both criteria and are omitted.
For two plate sources at $0.1\%$ reference noise, both selectors achieve 18/20 joint successes with $3\times8$ readings. With $4\times12$, the totals are 19/20 for OMP and 20/20 for LARS. For three sources, joint successes increase from 6/20 to 13/20 for OMP and from 3/20 to 14/20 for LARS.
At $1\%$ noise, individual-parameter recovery deteriorates
substantially even when the source count is correct.

The three-dimensional tests use the candidate grid and
tetrahedral sensor layout specified in Section~\ref{sec:protocol}.
At $0.1\%$ noise, LARS achieves 12/20 joint successes for two
sources and 3/20 for three with eight sampling times.
Increasing to sixteen times leaves the two-source result
unchanged and reduces the three-source total to 2/20.
Thus, additional temporal sampling produces no observed improvement in joint recovery for these configurations. The quantitative separation predictions
are assessed only in two dimensions.

Section~\ref{sec:verification} quantifies the plate-model
discrepancy and reports the forward-response and derivative checks.

\subsection{Numerical verification}\label{sec:verification}

Table~\ref{tab:verification} reports the convergence of the
finite-difference response to the method-of-images response,
together with image-truncation and derivative errors.
The response comparison uses unit sources at three locations,
three sensors, and eight logarithmically spaced times on
$[0.025,2]$, independently of the nested sampling sets used
in the reconstruction experiments. Response-matrix errors
are measured in the relative Frobenius norm.
The finite-difference error decreases consistently under
mesh refinement, while extending the image range from two
to three changes the response by only $2.35\times10^{-9}$.

For the finite-plate reconstruction trials, the maximum and
median relative model discrepancies are $1.83\times10^{-4}$
and $4.01\times10^{-5}$, respectively, measured in the
Euclidean observation norm relative to the finite-difference
data. The model-discrepancy norm is at most $0.1106\epsilon$,
where $\epsilon$ is the acceptance radius for the corresponding
observation design. The sufficient perturbation bound
$\norm{e}_2+\norm{\boldsymbol e_{\rm mod}}_2\leq\epsilon$
holds for 233 of the 240 data vectors; the maximum ratio
$(\norm{e}_2+\norm{\boldsymbol e_{\rm mod}}_2)/\epsilon$
is $1.0491$.

\begin{table}[htbp]
\centering\small
\caption{Finite-plate response convergence, image truncation,
and derivative verification.}
\label{tab:verification}
\begin{tabular}{lr}
\toprule
Check & Relative error \\
\midrule
FD, $65^2$ interior nodes & $4.69\times10^{-4}$ \\
FD, $129^2$ interior nodes & $1.27\times10^{-4}$ \\
FD, $257^2$ interior nodes & $3.56\times10^{-5}$ \\
FD, $513^2$ interior nodes & $1.01\times10^{-5}$ \\
Image range 2 versus 3 & $2.35\times10^{-9}$ \\
Plate location Jacobian & $2.86\times10^{-10}$ \\
Variable-projection residual Jacobian & $4.37\times10^{-10}$ \\
Normalized-response derivative & $1.73\times10^{-10}$ \\
Boundary response (maximum absolute value) & $1.31\times10^{-9}$ \\
\bottomrule
\end{tabular}
\end{table}

Independent finite-difference comparisons confirm the accuracy
of the analytic location, variable-projection, and
normalized-response derivatives, with relative errors listed
in Table~\ref{tab:verification}. The truncated image response
also satisfies the homogeneous boundary condition to a
maximum absolute error of $1.31\times10^{-9}$.

\FloatBarrier
\section{Discussion and conclusions}\label{sec:conclusions}

This study examined the recovery of an unknown number of constant
positive point heat sources from temperature histories recorded by
a few sensors. Sparse dictionary selection provides initial source
locations, continuous constrained fitting estimates their positions
and strengths, and a noise-based discrepancy rule selects among the
attempted source counts. The results distinguish three recovery
objectives: detecting the need for additional sources, estimating
individual source parameters, and estimating total strength and the strength-weighted center.

The analysis explains why these objectives have different sensitivity
to source separation and measurement noise. For a close pair with nonzero projected curvature, the local
one-source residual scales quadratically with separation, whereas
the linearized response change along a compensating unit parameter
direction is bounded by a constant times the cube of separation. Thus, the data may distinguish a pair from a
single source while providing weak information about its individual
parameters. Total strength and the strength-weighted center can
remain accurately estimated, provided the merged-source response
is locally well conditioned. The Gaussian approximation relates
local one-source rejection to noise and separation, without
guaranteeing accurate two-source reconstruction.

Paired numerical experiments support this distinction. Four sensors
allow accurate two-source recovery at low noise in the tested
two-dimensional configurations, while recovery becomes more
difficult as source count or noise increases. Additional sensors
improve some three-source results, whereas additional sampling
times alone do not consistently improve recovery. Unresolved pairs
can still yield accurate aggregate estimates. Comparisons with
random initialization show that sparse selection is not uniformly
superior, and sensitivity-based sensor designs do not consistently
improve joint parameter recovery. Independently generated
finite-plate data and three-dimensional tests further illustrate
the applicability and limitations of the procedure.

The present conclusions are based on synthetic data with known
thermal properties and noise levels. Further work should assess
experimental temperature measurements and uncertainties in
boundary conditions, material properties, and sensor response.
For practical thermal monitoring, the central implication is that
a satisfactory temperature fit should be accompanied by an
assessment of which source quantities the measurements can
reliably determine.

\begingroup
\small
\bibliographystyle{unsrtnat}
\bibliography{references}
\endgroup
\end{document}